\documentclass[10pt,journal,compsoc]{IEEEtran}

\usepackage{graphicx}
\usepackage{float}
\usepackage{url}
\usepackage{mathrsfs}
\usepackage{amsfonts}
\usepackage{color}
\usepackage{multirow}
\usepackage{graphicx} 
\usepackage{epstopdf}

\usepackage{amsmath}
\usepackage{subfigure}
\usepackage[square, comma, sort&compress, numbers]{natbib}
\usepackage{amssymb}

\usepackage{amsmath}

\ifCLASSINFOpdf

\else

\fi

\begin{document}

\title{Ethna: Analyzing the Underlying Peer-to-Peer Network of Ethereum Blockchain}

\author{\IEEEauthorblockN{Taotao Wang, \IEEEmembership{Member, IEEE}, Chonghe Zhao, Qing Yang, \IEEEmembership{Member, IEEE},
Shengli Zhang, \IEEEmembership{Senior Member, IEEE}, Soung Chang Liew, \IEEEmembership{Fellow, IEEE}}


\thanks{
	
	T. Wang, C. Zhao, Q. Yang and S. Zhang are with the College of Electronics and Information Engineering, Shenzhen University, Shenzhen 518060, China (e-mail: ttwang@szu.edu.cn; zhaochonghe\_szu@163.com; yang.qing@szu.edu.cn; zsl@szu.edu.cn)
	
	S. C. Liew is with the Department of Information Engineering, The Chinese University of Hong Kong (e-mail: soung@ie.cuhk.edu.hk).

}

}

\IEEEtitleabstractindextext{
\begin{abstract}
The peer-to-peer (P2P) network of blockchain used to transport its transactions and blocks has a high impact on the efficiency and security of the system. The P2P network topologies of popular blockchains such as Bitcoin and Ethereum, therefore, deserve our highest attention.  The current Ethereum blockchain explorers (e.g., Etherscan) focus on the tracking of block and transaction records but omit the characterization of the underlying P2P network. This work presents the Ethereum Network Analyzer (Ethna), a tool that probes and analyzes the P2P network of the Ethereum blockchain. Unlike Bitcoin that adopts an unstructured P2P network, Ethereum relies on the Kademlia DHT to manage its P2P network. Therefore, the existing analytical methods for Bitcoin-like P2P networks are not applicable to Ethereum. Ethna implements a novel method that accurately measures the degrees of Ethereum nodes. Furthermore, it incorporates an algorithm that derives the latency metrics of message propagation in the Ethereum P2P network. We ran Ethna on the Ethereum Mainnet and conducted extensive experiments to analyze the topological features of its P2P network. Our analysis shows that the Ethereum P2P network possesses a certain effect of small-world networks, and the degrees of nodes follow a power-law distribution that characterizes scale-free networks.
\end{abstract}
	
}

\maketitle

\IEEEpeerreviewmaketitle

\section{Introduction}
\IEEEPARstart{F}{IRST} introduced in Bitcoin by Satoshi Nakamoto, blockchain is a secure, verifiable and tamper-proof distributed ledger for supporting digital asset transactions \cite{nakamoto2008bitcoin}. With the ability to achieve consensus over a permissionless decentralized network \cite{garay2015bitcoin,pass2017analysis}, blockchain has become a disruptive technology in the fields of FinTech \cite{fanning2016blockchain}, Internet of Things (IoT) \cite{dai2019blockchain}, and supply chains \cite{abeyratne2016blockchain}. In 2014, Ethereum introduces smart contract to blockchain to fulfill various Turing-complete computing tasks in a decentralized manner \cite{buterin2014ethereum}. With smart contract, Ethereum greatly extends the application of blockchain by allowing the users to develop various decentralized applications (DApps). The next-generation blockchain will further boost the performance by adopting cutting-edge technologies such as novel consensus protocols, cross-chain methods, and sharding.

As the communication infrastructure, peer-to-peer (P2P) networks are a vital component of blockchain systems \cite{neudecker2018network}. The nodes of a blockchain send and receive messages containing transactions and blocks over a P2P network to achieve distributed consensus. The operational performance and stability of a blockchain system are affected by the message forwarding protocol, the peer discovery protocol and the topology of its underlying P2P network. For example, the propagation time of a block in the network is a key factor that affects the transaction processing capability of the blockchain \cite{neudecker2018network, decker2013information} --- the shorter the block propagation time, the higher the transaction processing capability of the blockchain. In addition, malicious nodes can exploit particular defects of the underlying P2P network to conduct various attacks on the blockchain, e.g., DDoS, Eclipse attack \cite{saad2020exploring, heilman2015eclipse}.	
	
Therefore, it is particularly important to analyze and understand the P2P networks of blockchain systems. However, the current Ethereum blockchain explorers (e.g., Etherscan \cite{Etherscan}) focus on the tracking of block and transaction records but omit the analysis and characterization of the underlying P2P network. Moreover, the existing analytical methods proposed for Bitcoin-like P2P networks \cite{decker2013information,donet2014bitcoin,miller2015discovering,neudecker2016timing,cao2020exploring} are not applicable to Ethereum, since Ethereum manages its P2P network using the Kademlia DHT structure that is fundamentally different from the unstructured P2P network adopted by the Bitcoin blockchain. Finally, many prior works on analyzing and measuring P2P networks \cite{dale2008evolution, su2013measurement, stutzbach2008characterizing, fauzie2011temporal, ripeanu2002mapping} focused on file-sharing systems, e.g., BitTorrent and Gnutella; these measurement methods cannot be applied to blockchain systems because they exploit protocols specific to the associated file-sharing systems (see the detailed discussion on related work in Section 2).

This work presents the Ethereum Network Analyzer (Ethna), a tool that probes and analyzes the P2P network of the Ethereum blockchain. Unlike other works \cite{venati2019counting,gencer2018decentralization,kim2018measuring,gao2019topology} that investigate the Ethereum P2P network by straightforwardly measuring some macro network  characteristics (e.g., the scale of the P2P network, the delay distribution among nodes and the geographical distribution of nodes), we measure and analyze some fine topological characteristics of the Ethereum P2P network. We first measure the degree distribution of Ethereum nodes by leveraging the random selection feature of the message forwarding protocol in the Ethereum P2P network (i.e., randomly selecting some neighbor nodes to forward messages). Since the randomness of message forwarding is closely related to the actual node degrees, our measured node degrees are accurate enough to reflect the characteristics of Ethereum network topology. In addition, we also exploit the message forwarding protocol of the Ethereum P2P network to analyze the transaction broadcast latency and further obtain the number of hops required to disseminate messages to the whole Ethereum P2P network. Based on the measured data and analytical results of Ethna, we obtain the following conclusion about the topological structures of the Ethereum P2P network:

\begin{itemize}
	
\item Most nodes in the Ethereum P2P network have degrees less than 50, the default maximum number of neighbor nodes allowed to connect with the node when starting the node; there are a few super nodes with very high degrees. The degree distribution of all the network nodes presents a power-law distribution, which characterizes scale-free networks.

\item The average delay to broadcast a transaction to the whole Ethereum P2P network is around 200 ms. It takes 3-4 hops to broadcast a new block or a new transaction to the whole network, indicating  that the Ethereum P2P network has a certain effect of small-world networks.

\end{itemize}	
The main contributions of this work are as follows:

\begin{itemize}
	\item We design a novel method that can accurately measure the degrees of the Ethereum nodes using a simple setup (specifically, we only need to deploy one node to probe the Ethereum P2P network) with feasible complexity (specifically, we only need to count the received messages at the deployed node).
	
	\item We propose an efficient algorithm to analyze the performance metrics of message propagation  in the Ethereum network, including the transaction broadcast latency and the number of hops required to  broadcast messages to the whole P2P network.
	
	\item We implement our measurement methods in Ethna using Go programming language and deploy this tool in Ethereum Mainnet. Our experiment results provide new insights into the P2P network of Ethereum that can help improve the network design of blockchain systems.
	
\end{itemize}
The rest of this paper is organized as follows. Section 2 reviews the related work. Section 3 presents background on the Ethereum P2P network. Section 4 introduces our network measurement method. Section 5 analyzes the measured data to derive the network topological features. Section 6 concludes this work.

\section{Related Work}

The P2P network of Bitcoin adopts an unstructured topology, the gossip message broadcast protocol and a random node discovery protocol \cite{neudecker2018network}. Work \cite{decker2013information} investigated the block and transaction propagation in the Bitcoin P2P network and found that forks on the Bitcoin blockchain are mainly determined by the message propagation latency. Work \cite{donet2014bitcoin} measured the size, the node geographic distribution, the stability, and the propagation delay of the Bitcoin P2P network.

Based on the adopted random node discovery protocol, some works \cite{miller2015discovering, neudecker2016timing} measured the degrees of the nodes in the Bitcoin P2P network. These results reveal that node degree distribution of the Bitcoin P2P network follows a power law, i.e., a few of the nodes have very large degrees and most of the nodes have very small degrees. The investigation in \cite{cao2020exploring} also measured the P2P network of Monero coin that has a network protocol similar to that of Bitcoin, and it found that the node degrees of Monero also follow a power-law distribution. Networks whose node degree distribution follows a power law are often called scale-free networks \cite{barabasi2013network}. Therefore, the measurement results of \cite{miller2015discovering, neudecker2016timing, cao2020exploring} can support the conclusion that the Bitcoin and Monero P2P networks resemble scale-free networks.

The current Ethereum blockchain explorers (e.g., Etherscan \cite{Etherscan}, Ethereum Blockchain Explorer \cite{etherchain} , and Ethereum Nodes Explorer \cite{ethernodes}) tracks the transaction, block and node records but omit the characterization of the underlying P2P network. Some characteristics of the Ethereum P2P network were studied \cite{venati2019counting,gencer2018decentralization,kim2018measuring}, including the scale of the Ethereum P2P network, the delay distribution among nodes and the geographical distribution of nodes. However, no conclusion was made on the Ethereum network topology. By far, the investigations on measuring the degrees of nodes in the Ethereum P2P network and analyzing the Ethereum P2P network topology based on the degree distribution are lacking.

The peer discovery protocol of Ethereum is quite different from those of Bitcoin and Monero coin, because the Ethereum P2P network adopts the K-bucket data structure in the Kademlia DHT protocol \cite{maymounkov2002kademlia} to discover network nodes and maintain node information. Thus, it turns out that to measure the degrees of network nodes in Ethereum and to study the Ethereum P2P network based on the distribution of node degrees are not straightforward. For example, \cite{gao2019topology} measured the degrees of peers in the Ethereum P2P network assuming the number of peers stored in K-buckets is the same as the peer degree; however, the peer degrees measured in \cite{gao2019topology} is far greater than the actual peer degree, since the actual node degree is often much smaller than the number of nodes stored in the K-buckets due to the leaving of nodes over time (we provide experiment results to support this point in Section 5).

Besides the measurement works on the P2P networks of blockchain systems, other existing works on measuring P2P networks mainly focused on file-sharing systems, e.g., BitTorrent and Gnutella. Most works first captured the topology of the P2P network and then measured the network features (e. g., node degrees, message propagation hops) from the captured topology \cite{ dale2008evolution, su2013measurement, stutzbach2008characterizing, fauzie2011temporal, ripeanu2002mapping}. The general idea to capture the P2P network topology of BitTorrent is to utilize the feature of its node discovery protocol, i.e., Peer Exchange (PEX) \cite{wu2010understanding}. According to PEX, when a new node is connected to one of the nodes in the network, the new node can further obtain the list of the neighbor nodes of this connected node, and then establishes new connections with the nodes in the list. In addition, nodes will periodically exchange information with their neighbor nodes about their newly connected and disconnected nodes \cite{wu2010understanding}. This feature of PEX was used to measure the P2P network topology of BitTorrent and analyze its P2P network in \cite{dale2008evolution, su2013measurement, stutzbach2008characterizing, fauzie2011temporal, ripeanu2002mapping}. However, in the P2P network of the Ethereum blockchain, nodes do not exchange the information of their neighbor nodes with each other for security reason, thus the measurement methods for the BitTorrent P2P network are not applicable to the Ethereum P2P network.

\section{Background}
Ethereum network communications are defined in three protocols, RLPx for node discovery and secure transport, DEVp2p for application session establishments, and the Ethereum subprotocol for application-level communications \cite{Eth22,Devp2p23,RLPX24} implemented in the Geth reference client of Ethereum \cite{VERSION29}. This section describes the message forwarding protocol adopted by the Ethereum subprotocol for propagating messages of transactions and blocks over the P2P network. Ethna exploits this protocol to derive useful information on the Ethereum P2P network. 

\subsection[A]{Functional Modules of Message Forwarding Protocol}
As shown in Fig. 1, message forwarding  at each Ethereum node is handled by several functional modules. These modules interact with each others to complete the whole message forwarding process. The {\tt P2P} module is responsible for communicating with the underlying P2P network: it exchanges messages with other neighbor nodes, delivers messages containing blocks and transactions to the protocol manager module, and keeps the messages specific to P2P network communication within the P2P module for processing (e.g., ping/pong messages). The protocol manager ({\ttfamily {\ttfamily ProtocolManager}}) module processes the received block and transaction messages and delivers them to the transaction pool ({\ttfamily TxPool}) and block processing ({\ttfamily {\ttfamily BlockProcess}}) modules respectively. The {\ttfamily TxPool} module is used to store the transactions that have not been recorded onto the blockchain. The {\ttfamily BlockProcess} module is used to process the blocks newly received from the network. 

The transactions in {\ttfamily TxPool} are arranged according to the accounts they belong to, as illustrated in Fig. 2. Each row in Fig. 2 arranges transactions issued by the same account, and these transactions are sorted by their nonce\footnote{Nonce in each transaction is an integer that is associated with the account that issues this transaction. For each account, the nonce value starts from 0 and it is increased by 1 after a new transaction is issued by this account.} values in ascending order. According to whether their nonce values are continuous, these transactions are respectively stored in two different subparts of {\ttfamily TxPool}:
\begin{figure}[t]
	\centering
	\includegraphics[width=3.5in]{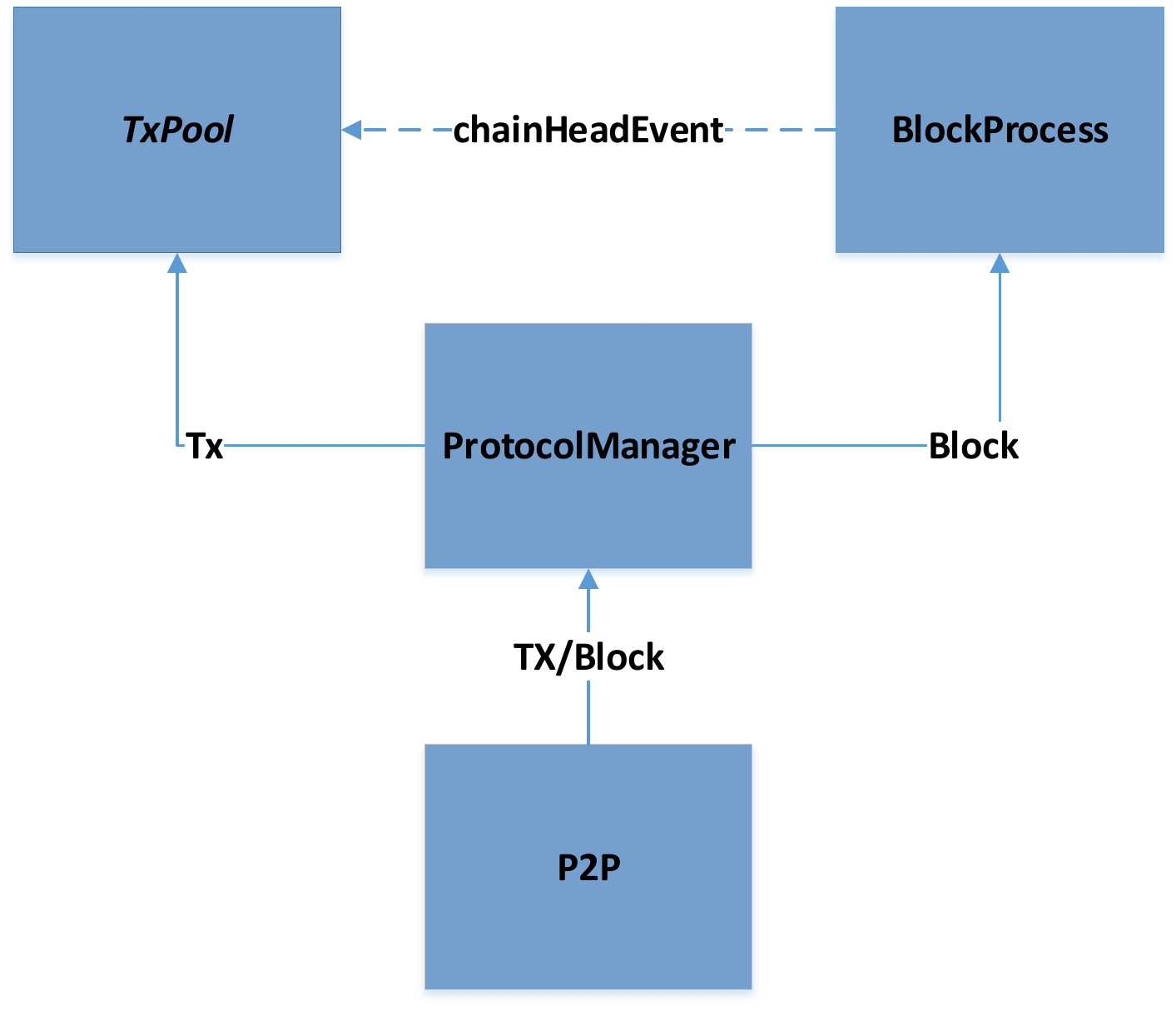}
	\caption{The functional modules of the message forwarding protocol adopted by the Ethereum P2P network.}
	\label{p2p_module}
\end{figure}

\begin{figure}[t]
	\centering
	\includegraphics[width=3.5in]{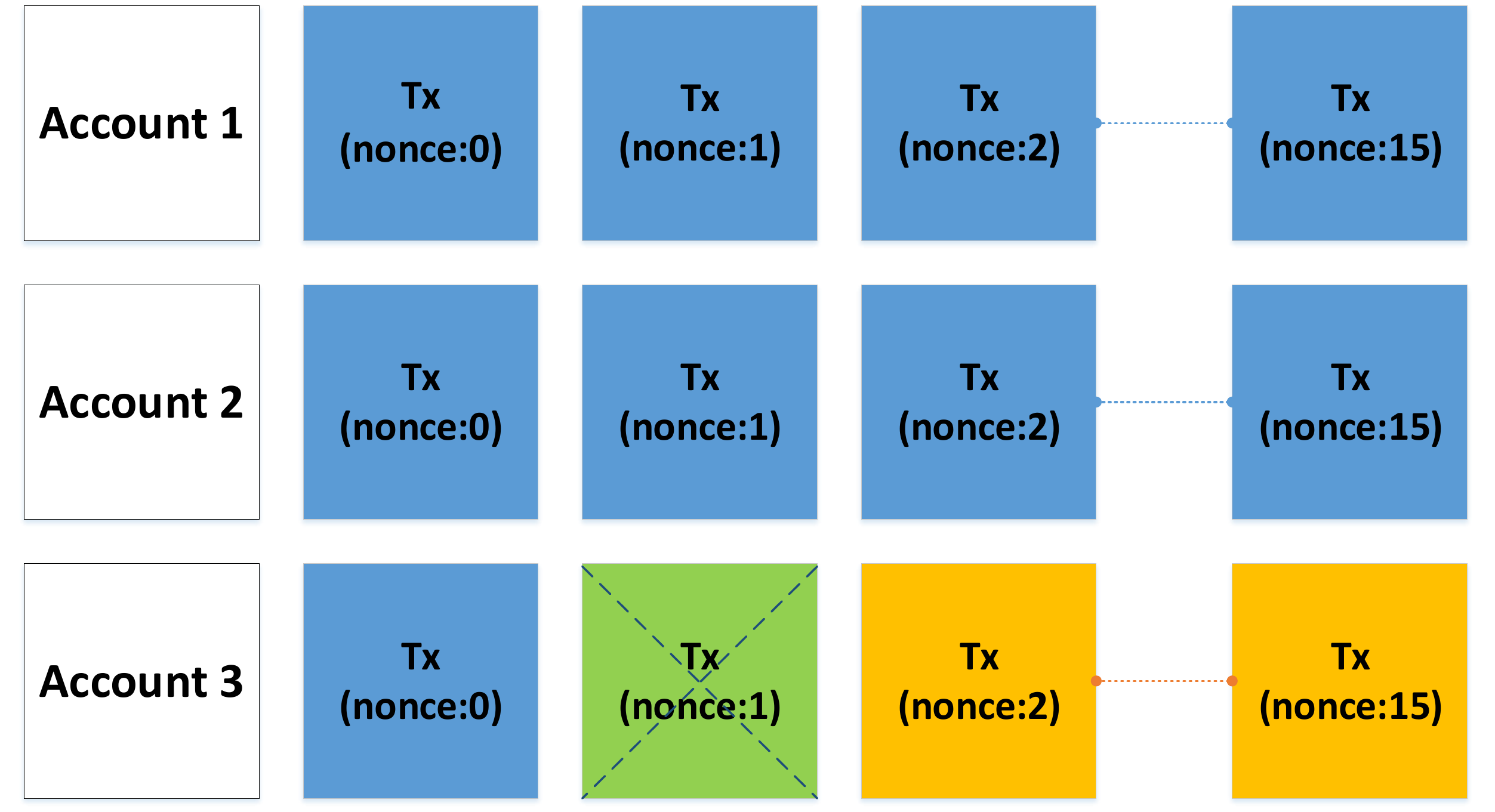}
	\caption{The arrangement of the transactions stored in {\ttfamily TxPool}.}
	\label{p2p_module}
\end{figure}

  \begin{itemize}
  	\item {\ttfamily PendingPool}: it maintains the pending transactions that have not been included in the blocks on the blockchain but are ready to be packaged into a new block. As shown in Fig. 2, the blue transactions are pending transactions; the nonce values of these transactions linked to each account are continuous. For each account, the maximal number of pending transactions can be stored in {\ttfamily PendingPool} is 16. In addition, once a transaction goes to the {\ttfamily PendingPool} of {\ttfamily TxPool}, the {\ttfamily TxPool} module will inform the {\ttfamily ProtocolManager} module that this transaction can be forwarded to other neighbor nodes who do not have the transaction yet. Miner nodes can select transactions from its {\ttfamily PendingPool} to pack them into a new block according to the packing rules of Ethereum.
  	\item {\ttfamily Queued}: it maintains the “future” transactions after a transaction with a continuous nonce value is absent and these future transactions are not ready to be packaged into a new block. As shown in Fig. 2, the transaction (in green) belonging to Account 3 with the nonce value of 1 is missing at the moment. Therefore, the later available consecutive transactions (in yellow) with nonce value greater than 1 enter the {\ttfamily Queued} of {\ttfamily TxPool}. Only after the transaction with nonce value of 1 arrives at {\ttfamily TxPool} can these transactions with continuous nonce values enter {\ttfamily PendingPool}. For each account, {\ttfamily Queued} can store up to 64 transactions.  	
  \end{itemize}
When a new block from the network arrives at the {\ttfamily ProtocolManager} module, it will be sent to the {\ttfamily BlockProcess} module for further processing. First, {\ttfamily BlockProcess} validates the block. Once the block passes validation, the world state of the local blockchain will be updated, and then a chain head event ({\tt ChainHeadEvent}) will be sent from {\ttfamily BlockProcess} to {\ttfamily TxPool}. {\tt ChainHeadEvent} contains the latest world state updated by the current new block. After receiving {\tt ChainHeadEvent}, {\ttfamily TxPool} will reset itself according to the updated world state. That is, it removes the transactions stored in {\ttfamily PendingPool}/{\ttfamily Queued} that are included in the new block; moreover, it adds the transactions included in the new block but not in the {\ttfamily PendingPool}/{\ttfamily Queued} to the {\ttfamily PendingPool}/{\ttfamily Queued}. During this period of the {\ttfamily TxPool} resetting, {\ttfamily TxPool} is temporarily blocked and the transactions sent from {\ttfamily ProtocolManager} cannot enter {\ttfamily TxPool}.

Normally, when a new block is received and there is no fork found on the blockchain, the transactions included in the new block will be removed from {\ttfamily PendingPool} and {\ttfamily Queued} during the {\ttfamily TxPool} resetting, as explained above. On the other hand, when two blocks with the same height are received in sequence, i.e., there is a fork observed on the blockchain, the processing of the {\ttfamily TxPool} resetting is different. The two blocks, which are denoted by {\tt FormerBlock} and {\tt AfterBlock}, respectively, correspond to two different world states. After receiving {\tt FormerBlock}, {\ttfamily TxPool} will be reset according to its world state, i.e., the transactions that are included in  {\tt FormerBlock} and also are existing in {\ttfamily PendingPool}/{\ttfamily Queued} are removed from {\ttfamily PendingPool}/{\ttfamily Queued}, and those only included in  {\tt FormerBlock} but are not existing in {\ttfamily PendingPool}/{\ttfamily Queued} are added to {\ttfamily PendingPool}/{\ttfamily Queued}. After receiving {\tt AfterBlock}, the transactions that are existing in {\tt AfterBlock} will be removed from {\ttfamily PendingPool}/{\ttfamily Queued}.  Then, the transactions in  {\tt FormerBlock} and {\tt AfterBlock} are compared to find the transactions that are not included in {\tt AfterBlock} but are included in  {\tt FormerBlock} (have already been excluded from {\ttfamily PendingPool}/{\ttfamily Queued}). These transactions need to be re-added to {\ttfamily PendingPool}/{\ttfamily Queued}.

From the above descriptions, we can see that when the {\ttfamily TxPool} resetting is happening, a large number of transactions may enter {\ttfamily PendingPool} simultaneously, and then {\ttfamily TxPool} informs {\ttfamily ProtocolManager} that these newly coming transactions can be forwarded to other nodes in the network.

\subsection[B]{Block Propagation Strategy}
In the Ethereum P2P network, the propagation of blocks and transactions adopt a gossip strategy \cite{aysal2009broadcast}. Let us  consider an Ethereum node that receives a new block from one of its neighbor nodes. The node that receives the new block will randomly select some of its connected neighbor nodes to propagate the received new block. We refer to the neighbor nodes of this node that currently do not know the new block as the downstream peers of the node. The number of these downstream peers is denoted by $N$. After the node receives the new block, it first validates the block header, and then it randomly selects $\sqrt N $ downstream peers to forward the new block.\footnote{Indeed, the number of neighbor nodes to propagate new messages determines a trade-off between the message broadcast complexity and the robustness/reliability of the gossip broadcast protocol. Generally speaking, large numbers of neighbor nodes to propagate increase the probability of reaching all nodes, but also generate more redundant traffic in the P2P network. Theoretical analysis of gossip-based message broadcast protocols which relates their reliability to the message broadcast complexity can be found in \cite{aysal2009broadcast, georgiou2008complexity}.} After that, it further validates the whole block, and then sends the hash of the block to the remaining $N - \sqrt N $ downstream peers if the block validation passes. Fig. 3 illustrates this block propagation process after an Ethereum node receives a new block.
\begin{figure}[t]
	\centering
	\includegraphics[width=3.5in]{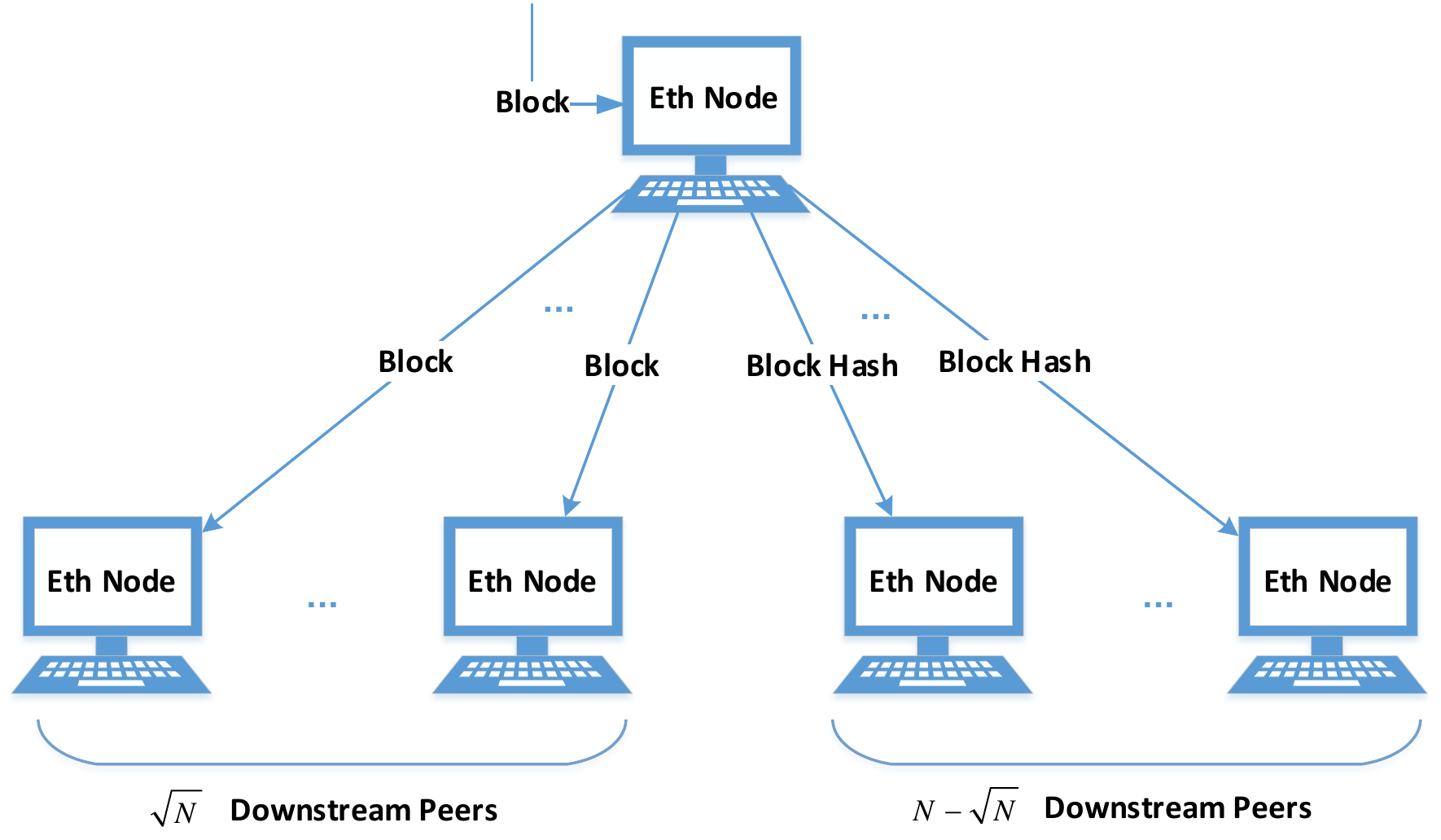}
	\caption{The block propagation process after an Ethereum node receives a new block.}
	\label{p2p_module}
\end{figure}

We next explain how an Ethereum node acquires a new block when the node receives the block hash from some of its neighbor nodes. Fig. 4 illustrates the process of acquiring a new block at an Ethereum node that receives a block hash. When an Ethereum node receives the new block hash, this node first waits for 400 ms; and then randomly selects one node from the neighbor nodes that already know the new block (i.e., the neighbor nodes that have already sent the block or the block hash to this node). After that, this node sends GetHeader information to request the header of the new block from the selected neighbor node. After receiving the block header returned by the selected neighbor node, the node will wait for 100 ms, and then randomly select a neighbor node from the set of the neighbor nodes that know the new block to obtain the body of the new block. In the end, the received new block body and the new block header will be assembled into a new block and appended to the tail of the local blockchain after the validation of this block is passed. Usually, an Ethereum node connects with multiple neighbor nodes in the Ethereum P2P network. As shown in Fig. 4, it is a long process for a node to obtain the block through the hash of this block. As a consequence, during the process, it is possible that other neighbor node may send the new block to this node. Once the node receives the block sent from other neighbor node, it will stop the process of obtaining the block through the block hash.

\begin{figure}[t]
	\centering
	\includegraphics[width=3.5in]{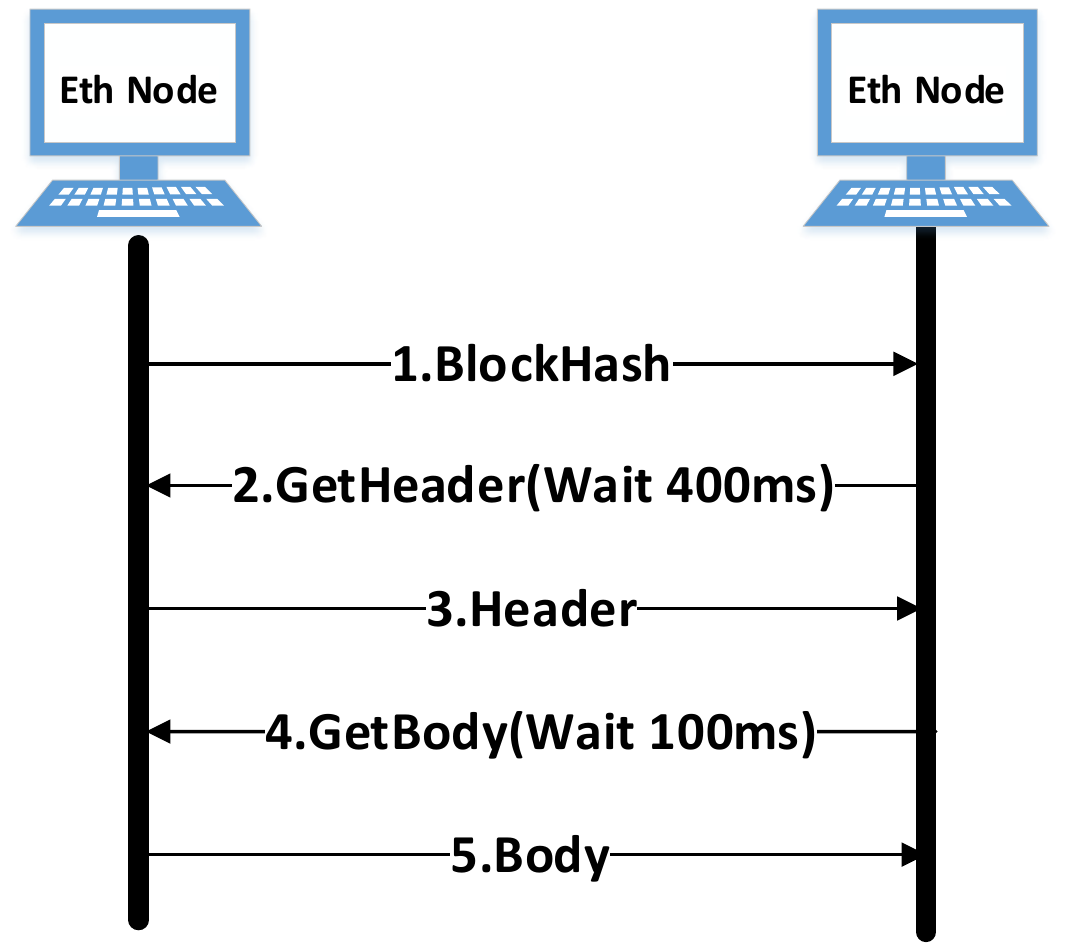}
	\caption{The block acquiring process at an Ethereum node after it receives a block hash.}
	\label{p2p_module}
\end{figure}

\subsection[C]{Transaction Propagation Strategy}
The nodes in the current Ethereum Mainnet run different protocols, such as the les protocol which provides light node service and the eth protocol that provides full node service \cite{LES26}. For  nodes running the les protocol, only the block header information will be synchronized; the transaction information in block body will be synchronized from other nodes when it is necessary. For the nodes running the eth protocol, all block information including transactions will be synchronized all the time. We aim to exploit the transaction propagation process to measure the Ethereum P2P network. Thus, we will focus on the nodes running the eth protocol.

The eth protocol has different versions, such as eth62, eth63, eth64 and eth65 \cite{ETH27}. The nodes running different versions of the eth protocol have slightly different transaction propagation strategies. Note that the transaction propagation strategy of eth64 and its earlier versions are completely different from that of eth65 and its later versions. Moreover, most of the eth protocols currently used by the nodes in Ethereum Mainnet are eth64 or eth65. Therefore, the transaction propagation strategies of eth64 and eth65 will affect our measurement method. In the following, we call the nodes running the eth64 protocol eth64 nodes, and the node running the eth65 protocol eth65 nodes. The transaction propagation strategies of eth64 and eth65 are described below.

For an eth64 node, receiving a new transaction, the {\ttfamily ProtocolManager} module sends the new transaction to {\ttfamily TxPool} for validation. The validated new transactions are stored into the {\ttfamily PendingPool} or {\ttfamily Queued} following the rules in Section 3.1. When a new transaction enters the {\ttfamily PendingPool} of the {\ttfamily TxPool} module, the {\ttfamily TxPool} module will inform the {\ttfamily ProtocolManager} module that a new transaction to forward to other nodes is available. The {\ttfamily ProtocolManager} module then forwards the new transaction to the neighbor nodes who currently do not have the transaction.

For an eth65 node, the transaction propagation processing executes two different actions according to the version of the eth protocol adopted by the forwarding target neighbor node: forwarding the transaction itself or forwarding the transaction hash to the target neighbor node. We illustrate the transaction propagation process of an eth65 node in Fig. 5. When an eth65 node receives a new transaction, its {\ttfamily ProtocolManager} module first sends the new transaction to {\ttfamily TxPool} for validation. When the transaction passes validation and is stored in the {\ttfamily PendingPool} of {\ttfamily TxPool}, {\ttfamily TxPool} notifies the {\ttfamily ProtocolManager} module that there is a new transaction that can be forwarded to other neighbor nodes. Then, the {\ttfamily ProtocolManager} randomly selects $\sqrt N $ downstream peers that do not know the transaction as the targets to forward this transaction. For the remaining $N - \sqrt N $ downstream peers, if the version of the eth protocol run by the downstream peer is eth65, the transaction hash will be forwarded; if the version of the eth protocol run by the downstream peer is eth64, then the transaction will be forwarded.

As explained above, eth65 nodes will receive transaction hashes from its neighbor eth65 nodes. When an eth65 node receives the hash of a new transaction, the process of obtaining the new transaction is illustrated in Fig. 6. After receiving the new transaction hash, the eth65 node waits for 500 ms. During this period, if no other neighbor node sends the new transaction to it, the eth65 node randomly selects one of the neighbor nodes that have sent it the new transaction hash, and sends a GetTx information to the selected neighbor node to request the new transaction. After the requested neighbor node returns the new transaction, the eth65 node validates the new transaction. If the transaction passes validation, the transaction is added to {\ttfamily TxPool} of this eth65 node.

\begin{figure}[t]
	\centering
	\includegraphics[width=3.5in]{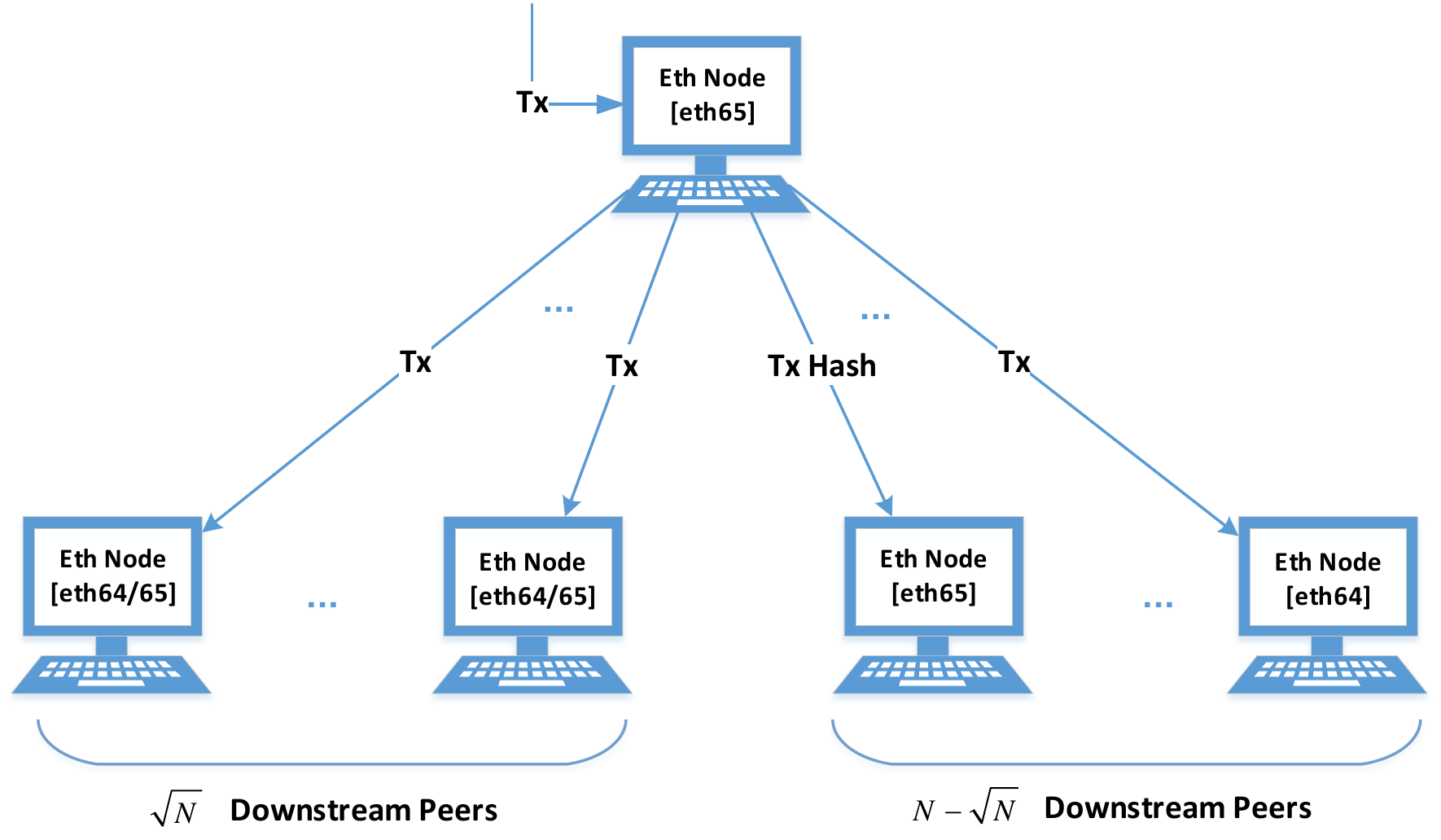}
	\caption{The transaction propagation process after an eth65 node receives a new transaction.}
	\label{p2p_module}
\end{figure}

\begin{figure}[t]
	\centering
	\includegraphics[width=3.5in]{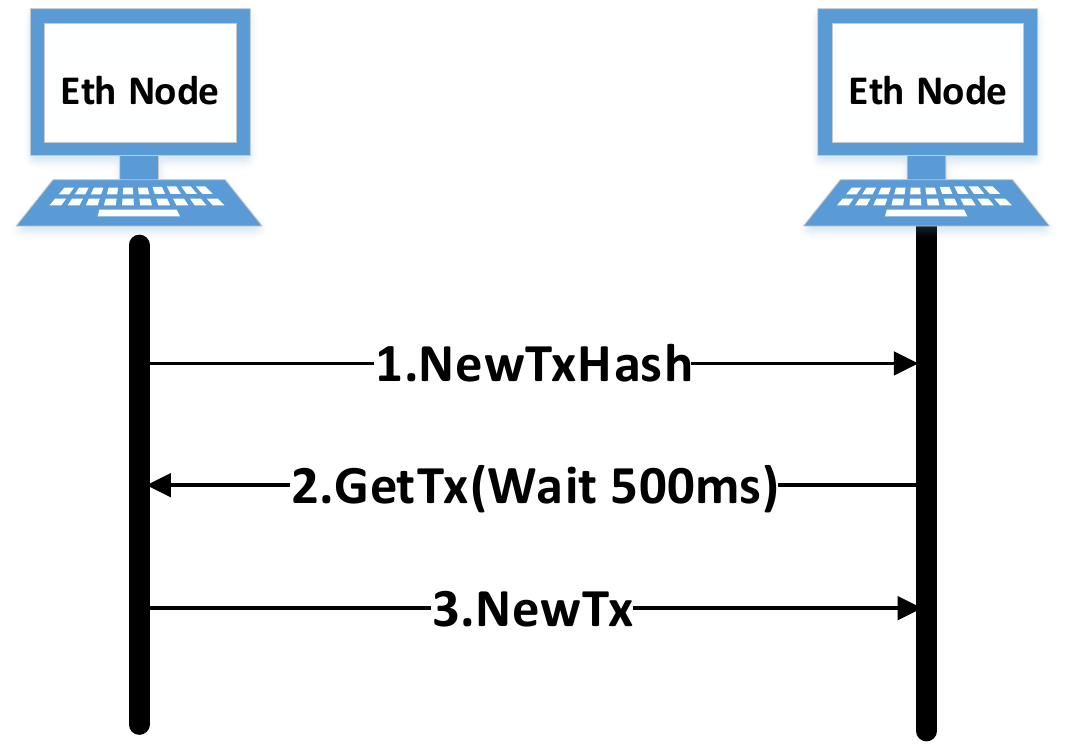}
	\caption{The process of obtaining a new transaction after an eth65 node receives the hash of the new transaction.}
	\label{p2p_module}
\end{figure}

\section{Network Measurement}
In this section, we introduce the Ethereum network nodes that are set up by Ethna to probe the Ethereum P2P network. Then, using these probing nodes, we propose methods to measure the time instants when nodes propagate transactions, and the numbers of transactions propagated by other Ethereum nodes. In our measurements, we assume that the network nodes follow the specific protocol behaviors as described in Section 3, which are implemented in the Ethereum's Geth reference client.\footnote{As shown in \cite{kim2018measuring}, the Geth clients account for around 76\% of the nodes in the Ethereum Mainnet, and thus this is not a strong assumption. Moreover, our measurement results obtained from the actual data of Ethereum Mainnet have verified the effectiveness of our measurement method.}

\subsection[A]{Setting up Probing Nodes}
To probe the Ethereum P2P network, we set up two different nodes on Ethereum Mainnet:
\begin{itemize}
	\item {\ttfamily NetworkObserverNode}: this is an Ethereum node that operates under the fast-synchronization mode. When the state of the local blockchain of a node is far from the world state of the current blockchain, the node will execute the fast-synchronization mode to rapidly synchronize to the current world state. The node operating under the fast-synchronizing mode only validate the world states contained in the blocks downloaded from other nodes and skips the validating and forwarding of the transactions included in the blocks. Thus, the fast-synchronization mode could reduce the synchronization time \cite{FAST_SYNC28}. {\ttfamily NetworkObserverNode} under the fast-synchronization mode will randomly connect with other neighbor nodes on Ethereum Mainnet. Although its neighbor nodes will send {\ttfamily NetworkObserverNode} new blocks and new transactions, {\ttfamily NetworkObserverNode} under the fast-synchronizing mode will not forward those blocks and transactions after receiving them. Therefore, {\ttfamily NetworkObserverNode} only plays the role of an observer on Ethereum Mainnet. {\ttfamily NetworkObserverNode} runs the Ethereum software with version v1.9.15 \cite{VERSION29} that adopts the eth65 protocol. We set up {\ttfamily NetworkObserverNode} using an AliCloud server located in Shenzhen, China, whose IP address over the public internet is 8.129.212.167.
	\item {\ttfamily LocalFullNode}: this is an Ethereum node that has synchronized to the latest word state of the blockchain. {\ttfamily LocalFullNode} will receive, verify and forward new blocks and new transactions. {\ttfamily LocalFullNode} runs the Ethereum software with version v1.9.15 that adopts the eth65 protocol. We set up {\ttfamily LocalFullNode} over the LAN of Shenzhen University in Shenzhen, China. {\ttfamily LocalFullNode} does not have its own public internet IP address, and it uses the NAT protocol to communicate with other nodes on Ethereum Mainnet.	
\end{itemize}

With these two Ethereum nodes, we can probe the Ethereum P2P network to measure transaction-propagation time instants, and the number of transactions forwarded by each node within a certain observation time window, as explained in the following.
\subsection[B]{Network Measuring Method}
\noindent {\bfseries1) Measuring transaction-propagation time instants:}

As discussed in Section 3, when Ethereum nodes join the network, they will choose suitable nodes to connect with and then exchange blocks and transactions with these connected neighbor nodes. Finding the time to broadcast a new transaction to most nodes in the Ethereum P2P network is one of the key objectives of Ethna. To achieve this objective, we first measure the transaction-propagation time instants at the neighbor nodes of {\ttfamily NetworkObserverNode} using the following measuring method. We utilize {\ttfamily NetworkObserverNode} to collect transactions sent from its connected neighbor nodes. We know that acting as a network observer, {\ttfamily NetworkObserverNode} will not change the states of {\ttfamily TxPool} at each of its neighbor nodes, because it only receives transactions but will not forward transactions. \footnote{Since the local blockchain state of {\ttfamily NetworkObserverNode} configured to execute the fast synchronization is far from the current world state of the blockchain, the recently issued transactions in Ethereum do not conform to the local blockchain state {\ttfamily NetworkObserverNode} and {\ttfamily NetworkObserverNode} will discard these new transactions when it fails to validate the transactions during the fast-synchronization process.}

Usually, a transaction or a transaction hash is propagated over the network in a packet solely consisting of this transaction or this transaction hash. For some cases, a number of transactions or transaction hashes will be encapsulated into one packet and propagated over the network.\footnote{A eth65 node will create two cache queues for each of its neighbor eth65 nodes, namely {\ttfamily TxQueued} and {\ttfamily TxHashQueued}. {\ttfamily TxQueued} is used to cache the corresponding transactions that are ready to be propagated to this neighbor node, and {\ttfamily TxHashQueued} is used to cache the corresponding transaction hashes that are ready to be propagated to this neighbor node. After the eth65 node selects a neighbor eth65 node to propagate a transaction or a transaction hash, it will immediately feed the transaction or the transaction hash into the {\ttfamily TxQueued} or {\ttfamily TxHashQueued} of this neighbor node. When the thread resources for processing the neighbor node are free, it will package all the transactions in the {\ttfamily TxQueued} into a transaction packet, and forward this transaction packet to the neighbor node. In the same way, the transaction hash needs to be cached in {\ttfamily TxHashQueued} of the neighbor node first before forwarding. When the thread resources are free, all transaction hashes in the {\ttfamily TxHashQueued} are packaged into a transaction hash packet and then forwarded to the corresponding neighbor node.} Whenever {\ttfamily NetworkObserverNode} receives a packet of transactions or transaction hashes from a neighbor node, we record the useful information about these transactions into a database called {\ttfamily TxMsgPool}. For each transaction or transaction hash contained in each received packet, we first record a raw-data record, {\ttfamily tempTxMsg}, that is given by the following form: 

{\ttfamily tempTxMsg} \{ {\ttfamily  PeerID},{\ttfamily TxHash},{\ttfamily TimeStamp},{\ttfamily GasPrice}, {\ttfamily PacketSize} \}

\noindent where the fields are explained below: {\ttfamily PeerID} is the peer identification of the neighbor node who sends the packet of transactions/transaction hashes to {\ttfamily NetworkObserverNode}; {\ttfamily TxHash} is the transaction hash; {\ttfamily TimeStamp} is the local time stamp when {\ttfamily NetworkObserverNode} receives the packet of transactions/transaction hashes; {\ttfamily GasPrice} is the service charge of this transaction paid to the miner; {\ttfamily PacketSize} is the number of total transactions/transaction hashes contained in the received packet (e.g., when the neighbor nodes sends a transaction to {\ttfamily NetworkObserverNode} and this transaction is separately encapsulated into a packet, the value of PacketSize is 1).

\begin{figure}[t]
	\centering
	\includegraphics[width=3.5in]{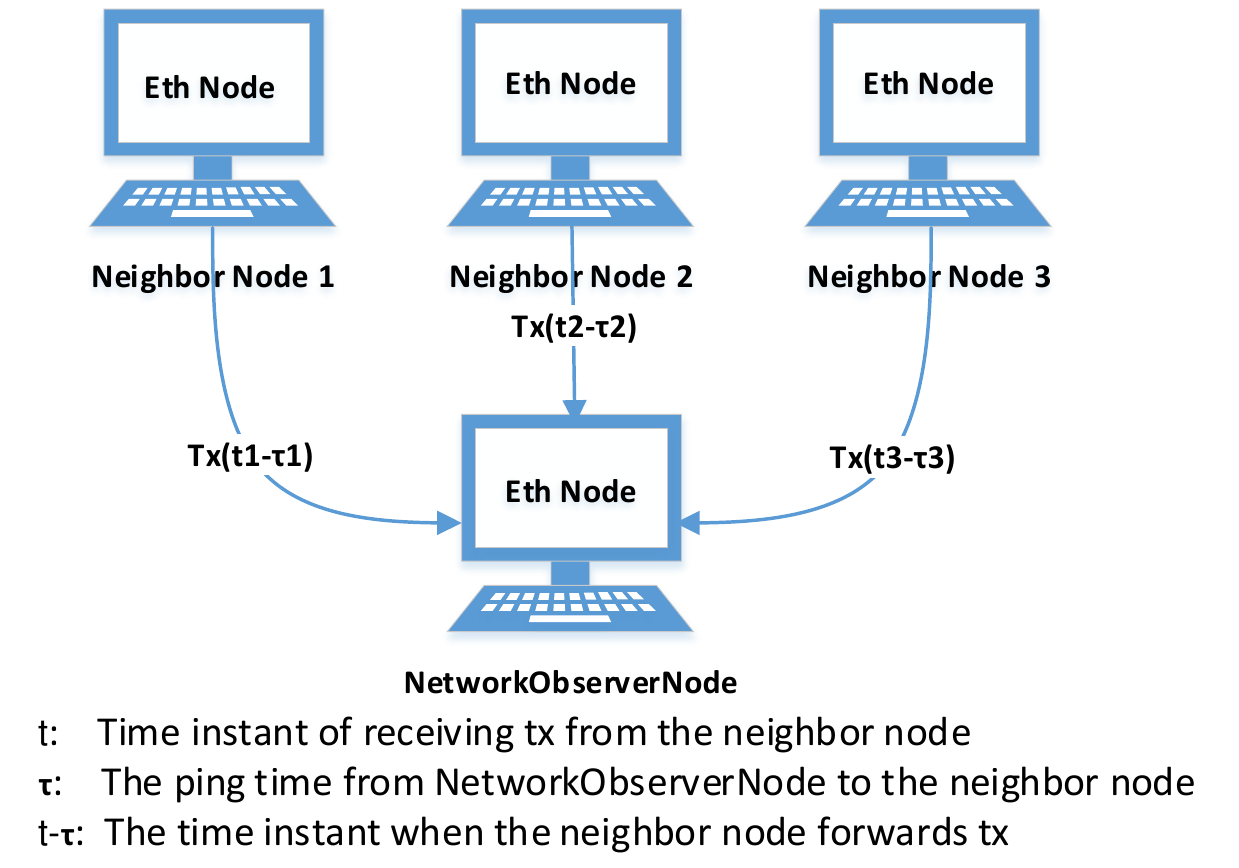}
	\caption{The measuring method for transaction propagation time instants between two neighbor nodes.}
	\label{p2p_module}
\end{figure}

\begin{table*}[htbp]
	\centering
	\caption{The impact of GasPrice on transaction propagations}
	\resizebox{\linewidth}{!}{
		\begin{tabular}{cccccc}
			\hline
			Location  & Received & Missed &\ {\tt GasPrice} $<$ 18 &\ 18 $\le $
			{\tt GasPrice} $\le $ 50 &\ {\tt GasPrice} $>$ 50 \\ \hline
			Beijing1 &1520& 8& 8& 0& 0 \\ \hline
			Beijing2 &1521& 7& 5& 0& 2\\ \hline
			Nuremberg &1505& 23& 23& 0& 0 \\ \hline
			Hangzhou &1528& 0& 0&0 & 0 \\ \hline
			Shenzhen1 &1519& 9& 9& 0& 0 \\ \hline
			Shenzhen2 &1506& 22& 22& 0& 0 \\ \hline
			Shijiazhuang &1503& 25& 23& 0& 2 \\ \hline
			Lille &1521& 7& 7& 0& 0 \\ \hline
			
		\end{tabular}
	}
\end{table*}

All raw-data records, {\ttfamily tempTxMsg}, are stored into the database, {\ttfamily TxMsgPool}. The TimeStamp filed in each raw-data record {\ttfamily tempTxMsg} is the local time stamp when {\ttfamily NetworkObserverNode} receives the transaction/transaction hash from a neighbor node. We also want  to measure the time instant at which this neighbor node forwards this transaction/transaction hash to {\ttfamily NetworkObserverNode}. Fig. 7 illustrates the measuring method. We denote the time cost of propagating a message from the neighbor node to {\ttfamily NetworkObserverNode} by {\tt Delay}, and  {\tt NetworkObserverNode} can measure the value of Delay by sending an ICMP-based ping packet to the corresponding neighbor node and taking half of the ping time between {\tt NetworkObserverNode} and the neighbor node.\footnote{TCP-based ping or the protocol's own ping/pong messages may be heavily influenced by network congestion, while ICMP-based ping often bypasses full queues and therefore tends to be more reliable.} The subtraction of {\tt Delay} from TimeStamp, i.e., {\tt TimeStamp} $-$ {\tt Delay}, gives the time instant at which the neighbor node forwards this transaction/transaction hash to {\ttfamily NetworkObserverNode}. However, when we measure the transaction-propagation time instants from the {\ttfamily tempTxMsg} records, we need to filter the {\ttfamily tempTxMsg}\{{\ttfamily PeerID}, {\tt TxHash}, {\tt TimeStamp}, {\tt GasPrice}, {\tt PacketSize}\} according to the following two criteria:
\begin{itemize}
	
	\item {\bfseries The GasPrice field of {\ttfamily tempTxMsg} should be no less than 18 Gwei.} The value of {\tt GasPrice} is an indicator to indicate whether this transaction is appropriately propagated over the network.  When an Ethereum node starts with the Geth client, the minimal value of the {\tt GasPrice} that determines whether transactions can enter its {\ttfamily TxPool} can be set. Only transactions with {\tt GasPrice} values that are no less than the set value can enter {\ttfamily TxPool} and be forwarded later. If the {\tt GasPrice} of a newly received transaction is too small, it will be discarded by the node immediately. As a consequence, the propagation time of that transaction will be longer; even worse, some nodes may fail to receive the transaction. The default minimal value of {\tt GasPrice} in the current version of Ethereum software is 1 Gwei, and in the older versions it is set to be 18 Gwei. We performed a simple experiment to investigate the impact of {\tt GasPrice} on transaction propagations. We first crawled 1528 transactions from the website \cite{Etherscan} on October 14, 2019. These transactions were propagated over the Etheruem P2P network. We then collected the propagation results of these transactions at our {\ttfamily NetworkObserverNode}. There are eight neighbor nodes (in different geographic locations) connected to {\ttfamily NetworkObserverNode}. These neighbor nodes forward their received transactions to {\ttfamily NetworkObserverNode}. We can check at {\ttfamily NetworkObserverNode} to see whether all 1528 transactions are propagated from the eight neighbor nodes to {\ttfamily NetworkObserverNode}. The propagation results of the 1528 transactions are shown in TABLE I. We can see that the numbers of these transactions received by {\ttfamily NetworkObserverNode} from each of its neighbor nodes are all less than 1528. This indicates that some transactions got lost. To look into these lost transactions, we find that most of the lost transactions have {\tt GasPrice} less than 18 Gwei. Thus, we can conclude that these transactions are discarded because of their small {\tt GasPrice}. Hence, in order to investigate the transactions that are propagated appropriately, we only select the raw-data records, {\ttfamily tempTxMsg}, whose {\tt GasPrice} field is no less than 18 Gwei for analysis.

	 \item {\bfseries The {\tt PacketSize} field of {\ttfamily tempTxMsg} should have a value of 1.} The value of PacketSize is a key indicator to determine whether the {\ttfamily TxPool} of the neighbor node who sends the transactions/transaction hashes to {\ttfamily NetworkObserverNode} is reset during the process of this transaction/transaction hash propagation. According to Section 3, if  {\ttfamily TxPool} resetting does not occur at the neighbor node, the neighbor node can store this transaction into its {\ttfamily TxPool} and forwards the transaction itself or the hash of this transaction to {\ttfamily NetworkObserverNode} promptly after this neighbor node receives the transaction and finishes its validation. In this case, the time cost used to validate and store this transaction is very short (around 1 ms) and it can be regarded as negligible. However, if there is {\ttfamily TxPool} resetting at the neighbor node, many transactions enter the {\ttfamily TxPool} simultaneously and then these transactions or the hashes of these transactions are forwarded together in one packet. Therefore, the propagation time of transactions is severely enlarged by the {\ttfamily TxPool} resetting process that leads to {\tt PacketSize} $>$ 1, and we only select the raw-data records, {\ttfamily tempTxMsg}, whose PacketSize filed is 1 for analysis.

\end{itemize}

Thus, using each raw-data record {\ttfamily tempTxMsg}\{{\ttfamily PeerID}, {\tt TxHash}, {\tt TimeStamp}, {\tt GasPrice}, {\tt PacketSize}\} where {\tt PacketSize} $=$ 1 and {\tt GasPrice} $>$ 18 and the corresponding {\tt Delay}, we can obtain a transaction-propagation record, {\tt TxMsg}, for this transaction propagation. The form of the transaction-propagation record {\tt TxMsg} is given by

{\ttfamily TxMsg} \{ {\ttfamily  PeerID},{\ttfamily TxHash},{\ttfamily ForwardTime} \}

\noindent where {\ttfamily ForwardTime} $=$ {\ttfamily TimeStamp} $-$ {\ttfamily Delay}. All the transaction-propagation records, {\ttfamily TxMsg}, are stored into the database {\ttfamily TxMsgPool} for the following analysis.

\noindent {\bfseries2) Measuring the number of transactions forwarded by each eth65 node:}

According to Section 3, when an eth65 node propagates a new transaction, it will forward both of the transaction and the transaction hash to some of its neighbor nodes. {Based on the transactions and transaction hashes received by {\tt NetworkObserverNode} during a certain measurement period, we can construct a node-propagation record, denoted by {\tt PeerPacketMsg}, for each node that connects with {\tt NetworkObserverNode}. The node-propagation record, {\tt PeerPacketMsg}, constructed for a node connecting with {\tt NetworkObserverNode}, stores the numbers of the transactions and the transaction hashes received by {\tt NetworkObserverNode} from this node.This node-propagation record is written as

{\ttfamily  PeerPacketMsg} \{{\ttfamily  PeerID},{\ttfamily  TxHashPacketCount},\\
 {\ttfamily  TxPacketCount}, {\ttfamily  StartTime},{\ttfamily  CurrentTime} \}

\noindent  where {\ttfamily PeerID} is the node identification, {\ttfamily TxHashPacketCount} is the transaction hash packets sent by this node, {\ttfamily TxPacketCount} is the number of the transaction packets sent by this node,  {\ttfamily StartTime} is the time instant that the first transaction or transaction hash packet sent by the node is received at {\ttfamily NetworkObserverNode}, {\ttfamily CurrentTime} is the local time that the latest transaction or transaction hash packet sent by the node is received at {\ttfamily NetworkObserverNode}.

So far, we have set up probing nodes over Ethereum Mainnet to collect transaction-propagation records, TxMsg, and the records of the transactions forwarded by each node, {\ttfamily PeerPacketMsg} from the Ethereum P2P network. In next section, we utilize these propagation records to analyze the topological features of the Ethereum P2P network.

\section{Analysis of Network Topological Features}

This section presents algorithms used by Ethna to analyze the topological features of the Ethereum P2P network, i.e, the distribution of node degrees, the latency cost to broadcast a transaction to most of the nodes, and the number of hops required to broadcast a message over the Ethereum P2P network. We implement Ethna in the Go programming language and deploy it in the current Ethereum Mainnet to experimentally verify Ethna.

\subsection[A]{The analytical method for the node degree distribution}

We first propose a novel and simple method for analyzing the degree distribution of Ethereum nodes.

As described in Section 3, whenever an eth65 node receives a new transaction, it forwards the transaction to $\sqrt N $ neighbor nodes that are randomly selected from the $N$ downstream peers that do not have this transaction, and forwards the transaction hash to the remaining $N - \sqrt N $ downstream peers. When a node receives a new transaction from one of its neighbor nodes, it needs to validate the transaction before forwarding it to other neighbor nodes. During the period from validating to forwarding, it could happen that its other neighbor nodes also forward the same transaction to this node. As a result, these neighbor nodes that also forward the transaction will not be treated as the downstream peers when this node forwards the transaction. Therefore, strictly speaking, the number of the downstream peers of this node should be smaller than the degree of the node. Moreover, for different transactions, the downstream peers and the numbers of the downstream peers may be different. In the Ethereum P2P network, the average processing time of a transaction after it is received by a node and before it can be forwarded by this node is around 1 ms. Compared with the average transaction propagation time between two neighboring nodes of 200 ms (see our measured result provided in Section 5.2), this processing time is negligible. For each node, we treat the numbers of the downstream peers for all of its forwarded transactions as the same number $N$, and we approximate its node degree as $N + 1$, where the plus 1 is due to that for each transaction, there is one neighbor node transmitting the transaction to the node.

In Section 4, for each of its connected nodes, {\ttfamily NetworkObserverNode} has obtained a node-propagation record, {\ttfamily PeerPacketMsg}, which stores the number of the transaction packets forwarded by this node and the number of the transaction hash packets forwarded by this node. Note that when forwarding each transaction, a node will randomly select $\sqrt N $ neighbor node from the $N$  downstream peers to forward the transaction and the remaining $N - \sqrt N $ downstream peers to forward the transaction hash\footnote{ Whether to forward the transaction or the transaction hash to a peer is determined by the following procedure. There is a table called peerSet that lists the eth65 node’s neighbor peers who do not know this transactions (i.e., the downsteam peers of this eth65 node about this transactions). The eth65 node selects the first $\sqrt N $ nodes from its peerSet to forward the transaction. And for each of its received transactions, the ordering of the neighbor peers in its peerSet is random. Therefore, even for the same group of the neighbor peers, their order in peerSet will be different for different transactions. This means that each time, the $\sqrt N $ neighbor peers where the eth65 node forwards the transaction are selected in a random manner.}. Therefore, the ratio of the number of the transactions received by {\ttfamily NetworkObserverNode} from a node over the number of the transactions plus the transaction hashes received by {\ttfamily NetworkObserverNode} from the same node is $\frac{{\sqrt N }}{N}$.  Based on the above analysis, we can establish the following formula for each node using the {\ttfamily TxPacketCount} and {\ttfamily TxHashPacketCount} data contained in its corresponding {\ttfamily PeerPacketMsg} record:
\begin{equation}
{{\sqrt N } \over N} = {{{\rm{{\tt TxPacketCount}}}} \over {{\rm{{\tt TxPacketCount}}} + {\rm{{\tt TxHashPacketCount}}}}}
\end{equation}
which holds statistically  when the number of transactions and transaction hashes forwarded by this node is large. Using the data contained in {\ttfamily PeerPacketMsg} for each node, we can solve (1) to find the number of the downstream peers $N$ and use $N + 1$ to as the approximation of the degree of this node. The computation of $N$ from (1) is rather feasible, i.e., we only need to collect the data about numbers of the received transactions and transaction hashes and have a simple computation as $N = {\left( {\frac{{{\rm{TxPacketCount}} + {\rm{TxHashPacketCount}}}}{{{\rm{TxPacketCount}}}}} \right)^2}$. The setup of collecting data used in the computation is also very simple, i.e., we only need to have a node of {\ttfamily NetworkObserverNode} as the network observer on Ethereum Mainnet. Next, we will conduct experiments to verify whether this estimation  method of node degrees in our Ethna is accurate.

\noindent {\bfseries1) The experimental verification of estimated node degrees:}

We ran {\ttfamily NetworkObserverNode} and {\ttfamily LocalFullNode} on Ethereum Mainnet to conduct the experiments of measuring the degrees of Ethereum nodes. We only used eth65 nodes as the targets to measure the node degrees of the Ethereum P2P network but did not use eth64 nodes. The reason for not using eth64 nodes to measure node degrees is explained in Appendix. We employ the network measuring method proposed in Section 4 to count the packets of transactions and transaction hashes sent by each of the eth65 nodes that are stably connected with {\ttfamily NetworkObserverNode} to obtain the {\ttfamily PeerPacketMsg} records. With the data contained in {\ttfamily PeerPacketMsg} records, we can solve the formula in (1) to obtain $N$ and use $N + 1$ to approximate the node degree for each eth65 node. After that, we can derive the degree distribution of Ethereum nodes.

We first conducted experiments to verify the measure method of node degrees in our Ethna. Our {\ttfamily NetworkObserverNode} and {\ttfamily LocalFullNode} were connected to each other on Ethereum Mainnet {during two different measure periods, i.e., the period from July 04, 2020 to August 05, 2020 and the period from December 12, 2020 to December 28, 2020. Since {\ttfamily LocalFullNode} has completed the blockchain synchronization, it will propagate transactions and transaction hashes to {\ttfamily NetworkObserverNode}. Therefore, {\ttfamily NetworkObserverNode} is used to collect the transaction packets and the transaction hash packets forwarded by {\ttfamily LocalFullNode} to obtain the {\ttfamily PeerPacketMsg} record of {\ttfamily LocalFullNode}. Then, the number of the downstream peers of {\ttfamily LocalFullNode}, $N$, is calculated by using the formula expressed in (1). We measure the network and compute $N$ for {\ttfamily LocalFullNode} on a daily basis, i.e., each day, we recount the numbers of transaction/transaction hash packets to obtain {\ttfamily PeerPacketMsg} and recalculate $N$ for {\ttfamily LocalFullNode}. In addition, we also count the numbers of block packets and block hash packets and calculate $N$ by using them to replace the numbers of transaction packets and transaction hash packets in (1). We then treat $N+1$ as the measured degree of {\ttfamily LocalFullNode} in our Ethna. We also measured the degrees of {\ttfamily LocalFullNode} using the K-bucket based scheme proposed in \cite{gao2019topology}. Fig. 8 presents the measured degrees of {\ttfamily LocalFullNode} and its actual degrees for comparisons over the two different measure periods. It can be seen from Fig. 8 that the measured node degrees using the method of our Ethna are very close to the actual node degrees. The measured node degrees using the K-bucket based scheme \cite{gao2019topology} are far larger than the actual node degrees. The measured results using our Ethna method with the numbers of transaction packets are closer to the actual node degrees than the measured results using our Ethna method with the numbers of block packets (the reason is explained in Appendix). We can see that the mismatches between the measured node degrees using our Ethna method with transaction packets and the actual node degrees range over [2, 4]. Therefore, it is regarded quite accurate to use the value of $N+1$ measured by our Ethna to approximate the node degree. Based on the measured degrees of the nodes that are stably connected with our {\ttfamily NetworkObserverNode}, we then can infer the degree distribution of all the nodes in the Ethereum P2P network.

\noindent {\bfseries2) The experiment for deriving the degree distribution:}

With the measured node degrees, we can analyze the degree distribution of nodes in the Ethereum P2P network. The right way to analyze the topology of P2P networks, such as the degree distribution of nodes, should be performed by taking a snapshot of the states of all nodes in the whole P2P network at the same time. When the Ethereum P2P network operates stably, the degrees of its nodes vary within a narrow range without significant fluctuations. And the varying range is related to {\tt MaximumPeerCount} (the maximum number of neighbor nodes allowed to connect with when starting the node) and the local network configuration. We can observe the degrees of our {\ttfamily NetworkObserverNode} and {\ttfamily LocalFullNode} nodes to verify whether the degrees of Ethereum nodes are stable over time within an acceptable range.

\begin{figure}[!t]
	\centering
	\includegraphics[width=3.5in]{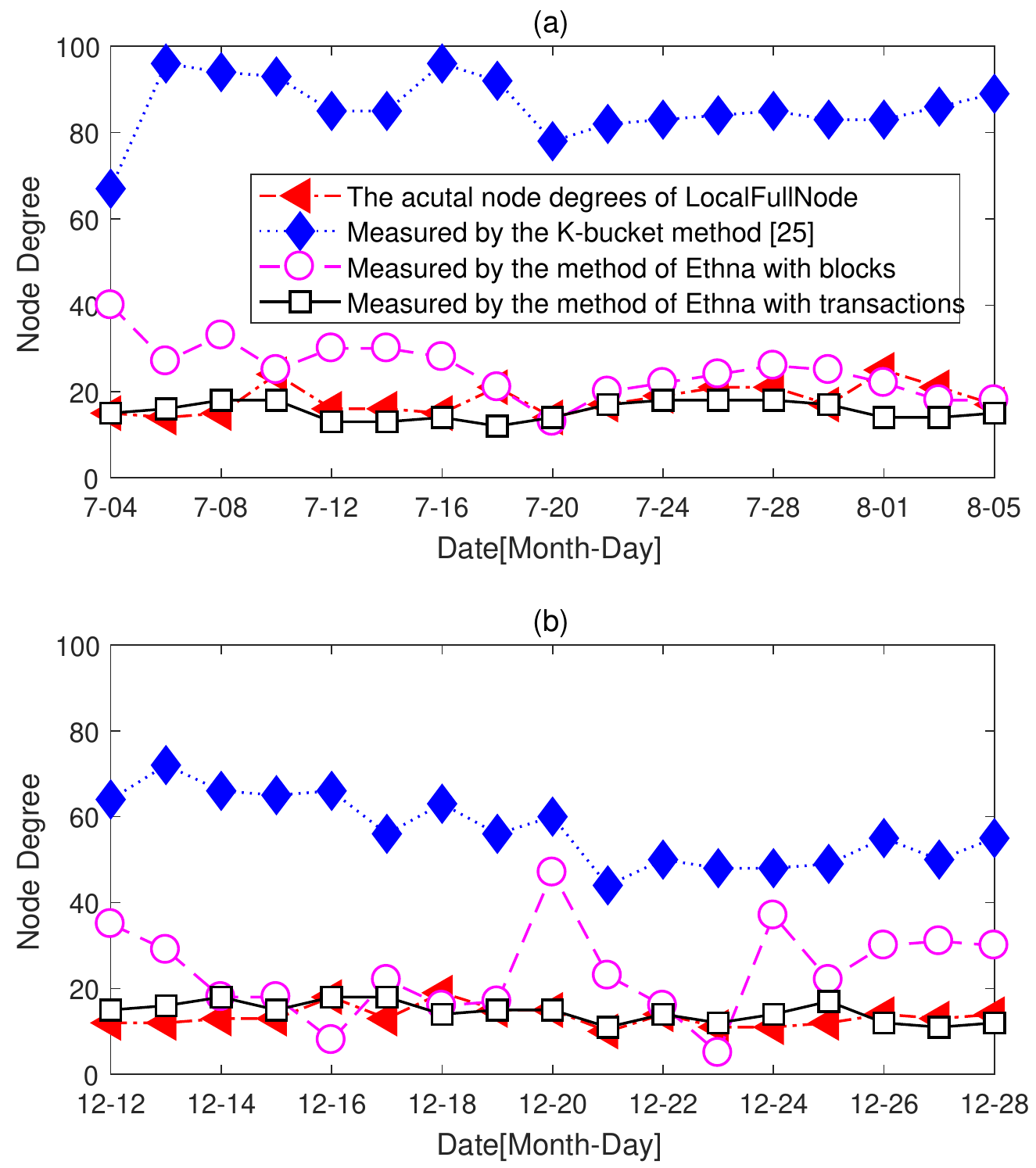}
	\caption{The comparisons between the measured degrees and the actual degrees of {\ttfamily LocalFullNode}: (a) from July 4, 2020 to August 05,2020; (b) from December 12, 2020 to December 28, 2020.}
	\label{p2p_module}
\end{figure}

Fig. 9 presents the observed degrees of {\ttfamily NetworkObserverNode} and {\ttfamily LocalFullNode} during the period from June 9, 2020 to June 17, 2020 and the period December 12, 2020 to December 20, 2020. The {\tt MaximumPeerCount} of the {\ttfamily NetworkObserverNode} and {\ttfamily LocalFullNode} is set to 50. From the results in Fig. 9, we can see that during the observation periods, the degree of {\ttfamily NetworkObserverNode} fluctuated within the range of [25, 30] and that of {\ttfamily LocalFullNode} fluctuated within the range of [10, 14]. This shows that the node degrees of {\ttfamily NetworkObserverNode} and {\ttfamily LocalFullNode} both are rather stable over different time. Therefore, we believe that when analyzing the degree distribution of Ethereum nodes, the degree of eth65 nodes measured at different time instants can be used to derive the degree distribution of the nodes in the Ethereum P2P network.

\begin{figure}[!t]
	\centering
	\includegraphics[width=3.5in]{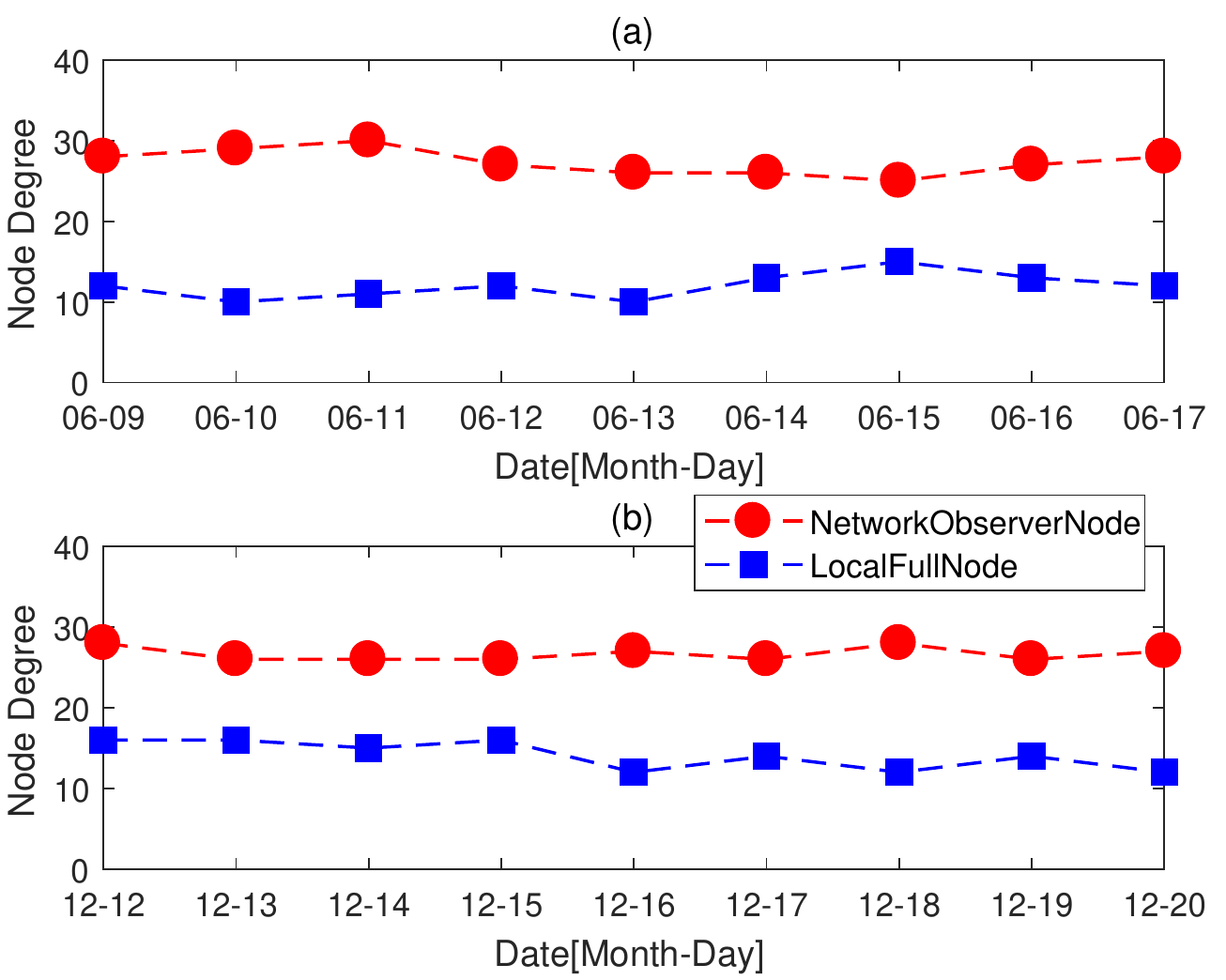}
	\caption{The node degrees of {\ttfamily NetworkObserverNode} and {\ttfamily LocalFullNode} observed during two different periods: (a) from June 9, 2020 to June 17, 2020; (b) from December 12, 2020 to December 20, 2020.}
	\label{p2p_module}
\end{figure}

We next run {\ttfamily NetworkObserverNode} to measure the degrees of the eth65 nodes that establish stable connections with {\ttfamily NetworkObserverNode} (if a node propagates more than 1000 packets of transactions or transaction hashes to {\ttfamily NetworkObserverNode}, it is regarded as a node that establishes a stable connection). We found that there were 555 eth65 nodes having stable connections with {\ttfamily NetworkObserverNode} during the measurement period from June 9, 2020 to June 17, 2020, and there were 769 eth65 nodes having stable connections with {\ttfamily NetworkObserverNode} during the measurement period from December 12, 2020 to December 20, 2020. We then analyze their degrees using the method of  Ethna. Fig.10 shows the empirical probability density functions (pdf) and Fig. 11 shows empirical cumulative distribution functions (cdf) of the node degrees, respectively. From Fig. 10 and Fig. 11, we can see that there are a small number of super nodes (nodes have very high degrees) in the network; the average degree of the 555 eth65 nodes is 47 during the period from June 9, 2020 to June 17 and the average degree of the 769 eth65 nodes is 64 during the period from December 12, 2020 to December 20, 2020. The version of Ethereum software is updated to v1.9.0 on June 12, 2020, which sets the default maximum degree of nodes to 50. From the results in Fig. 10 and Fig. 11, we can see that during the period from June 9, 2020 to June 17, the degrees of the 78\% of the nodes are smaller than {\tt MaximumPeerCount} and the degrees of 22\% of the nodes are greater than {\tt MaximumPeerCount}; during the period from December 12, 2020 to December 20, 2020, the degrees of the 70\% of the nodes are smaller than {\tt MaximumPeerCount}  and the degrees of 30\% of the nodes are greater than {\tt MaximumPeerCount}.} It reveals that most nodes started from {\tt MaximumPeerCount}, and only a few nodes modify {\tt MaximumPeerCount}  to act as super nodes.

For both of the measurement periods, the degrees of  the Ethereum network nodes obey a power-law distribution, whose pdf is given by $P\left( k \right) \sim {k^{ - \gamma }}$, where $k$ is the node degree, and $\gamma$ is the power law exponent of the function. By fitting our measured data into this pdf of power-law distribution, we find that the value of $\gamma$ for the Ethereum P2P network is 2.3363 during the period from June 9, 2020 to June 17, 2020, and it is 2.3765 during the period from December 12, 2020 to December 20, 2020. These values are very close to the value of $\gamma$ for the Bitcoin P2P network, which was measured to be 2.3 in \cite{neudecker2016timing}. Work \cite{clauset2009power} proposes to use the p-value of null hypothesis significance testing to verify whether the empirical distribution of collected data is power-law. The p-value is the evidence against a null hypothesis; the smaller the p-value, the stronger the evidence that you should reject the null hypothesis. Usually, a p-value that is less than the threshold value of 0.05 is regarded as a condition for the null hypothesis to be invalid. In our testing, the null hypothesis is that the empirical distribution of collected data is power-law. In \cite{clauset2009power}, the p-value threshold is raised from 0.05 to 0.1, i.e., it is believed that the hypothesis of power law distribution is not valid when the p-value of the data sample is less than 0.1. Based on the p-value method, we use our data collected from the Ethereum P2P network to fit the degree distribution of the nodes. We find that the p-value is 0.15 for the data collected during the period from June 9, 2020 to June 17, 2020 is 0.15 and it is 0.11 for the data collected during the period from December 12, 2020 to December 20, 2020. Therefore, the verification of the node degree distribution using p-value supports the conclusion that the node degrees of the Ethereum P2P network conform to the power law distribution. Since the node degree distribution of scale-free networks follows a power law \cite{barabasi2003scale, barabasi2013network}, we can conclude that the Ethereum P2P network resemble scale-free networks.

\begin{figure}[t]
	\centering
	\includegraphics[width=3.5in]{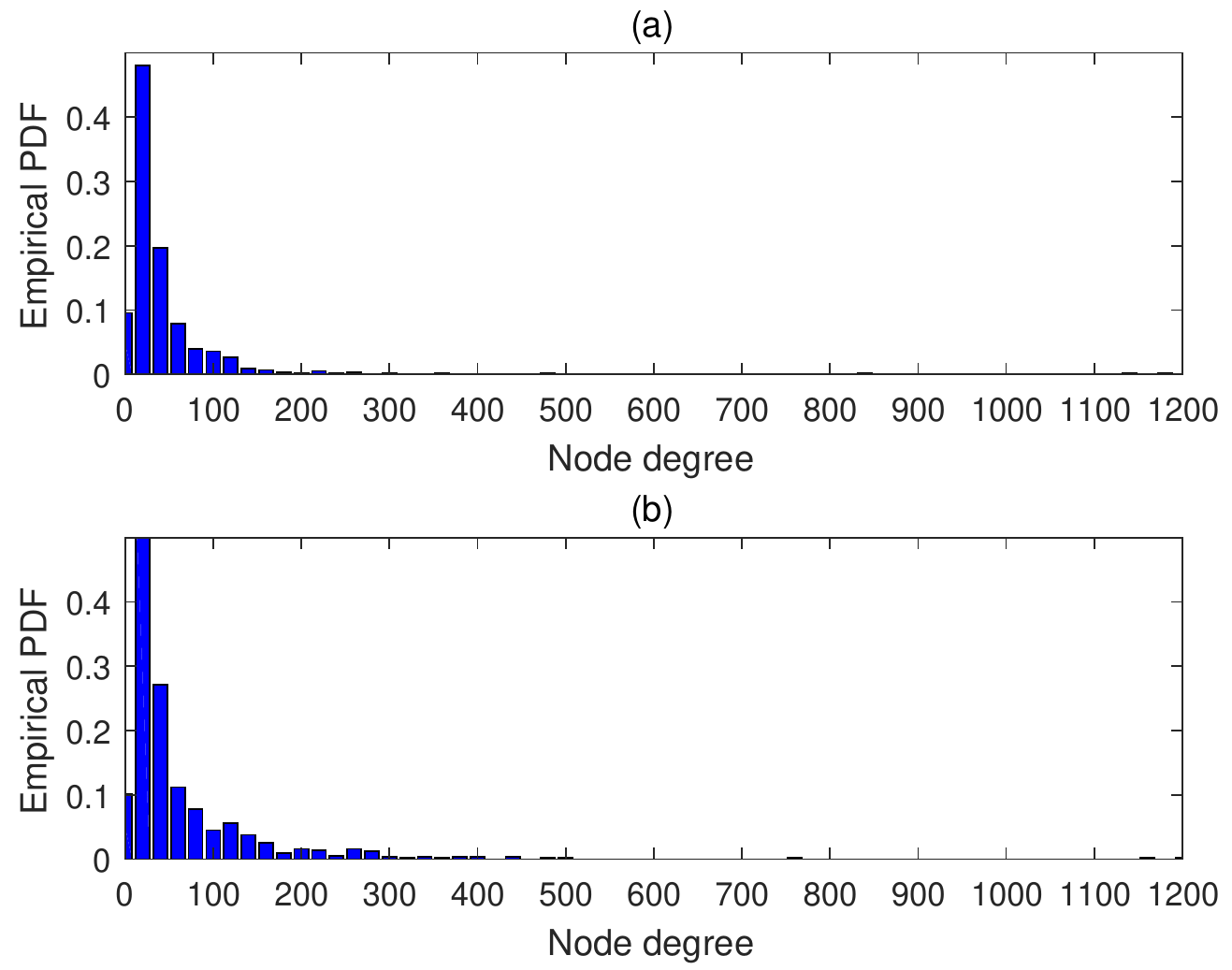}
	\caption{The empirical PDF of Ethereum eth65 Node Degrees during two different periods: (a) from June 9, 2020 to June 17, 2020; (b) from December 12, 2020 to December 20, 2020.}
	\label{p2p_module}
\end{figure}

\begin{figure}[t]
	\centering
	\includegraphics[width=3.5in]{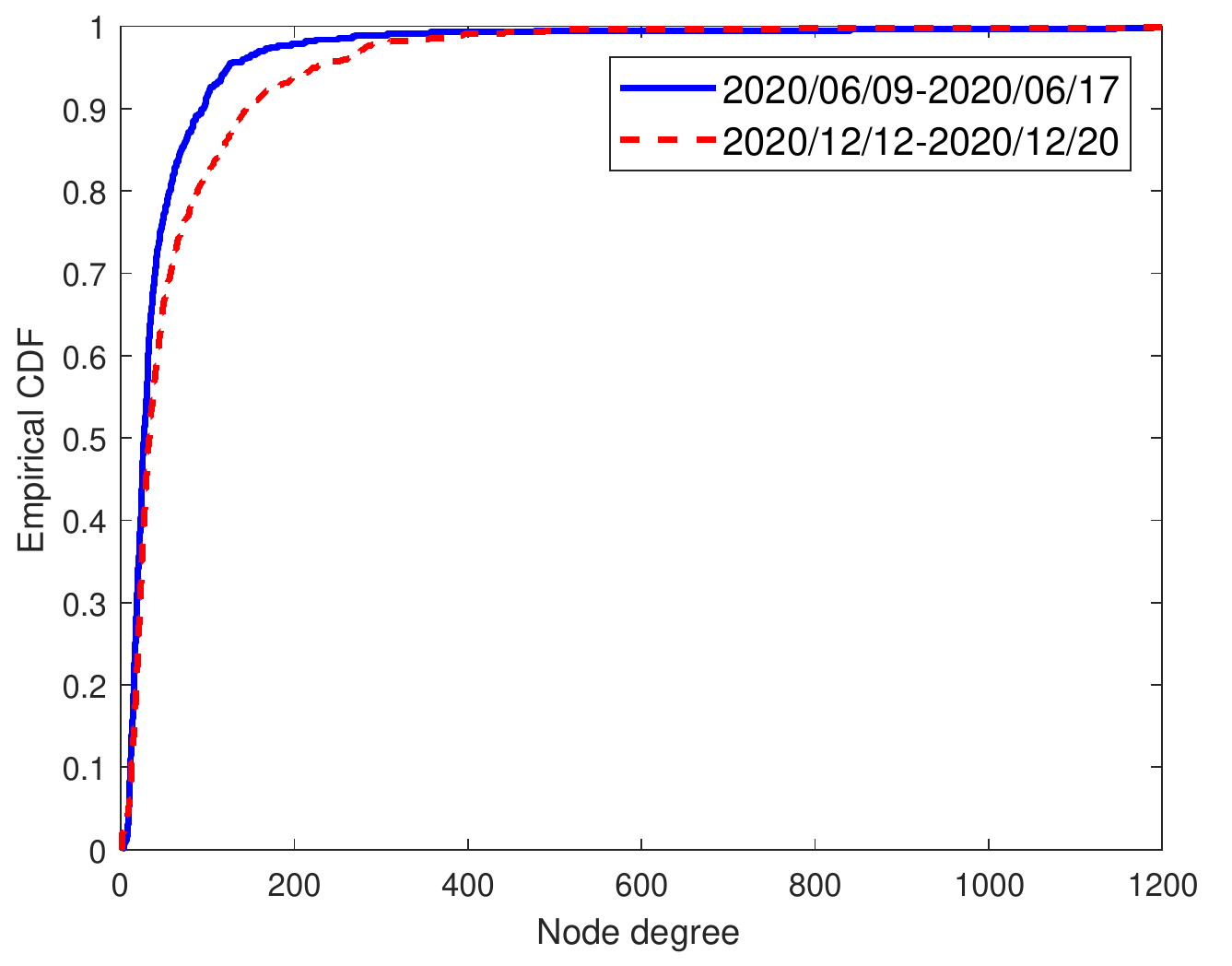}
	\caption{The empirical CDFs of Ethereum eth65 Node Degrees over two different periods: from June 9, 2020 to June 17, 2020 and from December 12, 2020 to December 20, 2020.}
	\label{p2p_module}
\end{figure}

\subsection[B]{The analytical method for transaction broadcast latency}

This part presents how Ethna analyzes the latency of broadcasting a transaction to most of the nodes in the Ethereum P2P network.  

According to Section 4.2, we write the propagation information of each transaction contained in each packet received by {\ttfamily NetworkObserverNode} into a transaction-propagation record, {\tt TxMsg}\{{\ttfamily PeerID}, {\tt TxHash}, {\tt ForwardTime}\}, and all {\tt TxMsg} records are stored in {\ttfamily TxMsgPool}. Since {\ttfamily NetworkObserverNode} is connected with multiple neighbor nodes at the same time, multiple nodes will forward the same transaction to {\ttfamily NetworkObserverNode}. Therefore, there will be more than one {\tt TxMsg} in {\ttfamily TxMsgPool} that corresponds to the same transaction, i.e., the TxHash fields of these {\tt TxMsg} records are the same. Therefore, we can extract all {\tt TxMsg} records corresponding to the same transaction from {\ttfamily TxMsgPool}, and select them to construct a set of the transaction-propagation records for the same transaction:

{\tt TxMsgSet} $=$ \{ {\tt TxMsg}[1],{\tt TxMsg}[2],..., {\tt TxMsg}[n]\}

\noindent where {\tt TxMsg}[$i$]\{{\ttfamily PeerID}[$i$], {\ttfamily TxHash}[$i$], {\ttfamily ForwardTime}[$i$]\} is the $i$-th transaction-propagation record for this transaction. In the set {\tt TxMsgSet}, the {\tt TxHash}[$i$] of all {\tt TxMsg}[$i$] are the same, and the {\tt ForwardTime}[$i$] of each {\tt TxMsg}[$i$] is the time instant that the corresponding neighbor node forwards this transaction to {\ttfamily NetworkObserverNode}. 

Based on the transaction-propagation records TxMsg in {\ttfamily TxMsgPool}, the set {\tt TxMsgSet} can be built for each transaction. With all the {\tt TxMsgSet} sets for the recorded transactions, we propose an algorithm to compute the latency to broadcast a transaction to most of the nodes in the Ethereum P2P network. The algorithm for computing the transaction broadcast latency is explained below:

\begin{enumerate}	
\item First, for each neighbor node that forwards transactions to {\ttfamily NetworkObserverNode}, we build an empty set called {\tt PeerTimeDiffSet}[{\ttfamily PeerID}], where {\ttfamily PeerID} is the network peer identification of the neighbor node.

\item We select a transaction and fetch the set of the transaction-propagation records, {\tt TxMsgSet}, for this selected transaction. Then we calculate the minimum value of the {\tt ForwardTime} fields of all the transaction-propagation records, {\tt TxMsg}, in {\tt TxMsgSet} and name this minimum value as {\tt minTime} that is the earliest time instant that this transaction is forwarded by a neighbor node to {\ttfamily NetworkObserverNode}.

\item For each and every {\tt TxMsg}[$i$]\{{\ttfamily PeerID}[$i$],{\tt TxHash}[$i$],{\tt ForwardTime}[$i$]\} in {\tt TxMsgSet}, we first calculate the difference between {\tt ForwardTime}[$i$] in {\tt TxMsg}[$i$] and {\tt minTime}, i.e., {\tt ForwardTime}[$i$] $-$ {\tt minTime}; then, we put the time difference, {\tt ForwardTime}[$i$] $-$ {\tt minTime}, into the corresponding set {\tt PeerTimeDiffSet}[{\ttfamily PeerID}[$i$]] according to the node’s peer identification, {\ttfamily PeerID}[$i$].

\item We repeat step 2) and step 3) for each and every recorded transaction to get the time difference sets, {\tt PeerTimeDiffSet}[{\ttfamily PeerID}], for all neighbor nodes that forward transactions to {\ttfamily NetworkObserverNode}.

\item We calculate the average value of all entries in the set of {\tt PeerTimeDiffSet}[{\ttfamily PeerID}] and denote this average value by {\tt PeerTimeDiffMean}[{\ttfamily PeerID}] for each neighbor node with peer identification, {\ttfamily PeerID}.

\item By averaging all {\tt PeerTimeDiffMean}[{\ttfamily PeerID}], we can get the estimated average latency to broadcast a transaction to most of the nodes in the Ethereum P2P network.

\end{enumerate}

\noindent {\bfseries1) The experiment for finding the transaction broadcast latency:}

We use the above algorithm with the collected transaction-propagation records to analyze the average transaction broadcast latency. However, when doing so, we need to ensure that we only use the propagation records of newly issued transactions that are first broadcasted over the network rather than some previous transactions that are repeatedly broadcasted over the network due to some problems. Therefore, it’s necessary to identify the new transactions of Ethereum in our analysis. The website \cite{Etherscan} publishes new transactions observed from Ethereum Mainnet in real time according to its historical records. Therefore, when running {\ttfamily NetworkObserverNode} to measure transaction propagation records, we use a crawler to collect the information about the new transactions published by the website. When we analyze the transaction broadcast latency, we only examine the propagation records of the new transactions.

To ensure that the transaction-propagation records collected by {\ttfamily NetworkObserverNode} can reflect the feature of the Ethereum network as much as possible, when the data analysis is conducted each time, we ensure that {\ttfamily NetworkObserverNode}  is connected with more than 20 neighbor nodes that are distributed all over the world. We conducted the network measurement on a daily basis from June 09, 2020 to June 17, 2020 over Ethereum Mainnet to get the  transaction-propagation records and estimate the average transaction broadcast latency within each day. Fig. 12 presents the results of the analyzed average transaction broadcast latency. We can see that the transaction broadcast latency of the Ethereum P2P network is relatively stable during the measuring period and it is slightly fluctuated around 200 ms. Therefore, we treat the average transaction broadcast latency of the Ethereum P2P network as 200 ms in our later analysis.

\begin{figure}[t]
	\centering
	\includegraphics[width=3.5in]{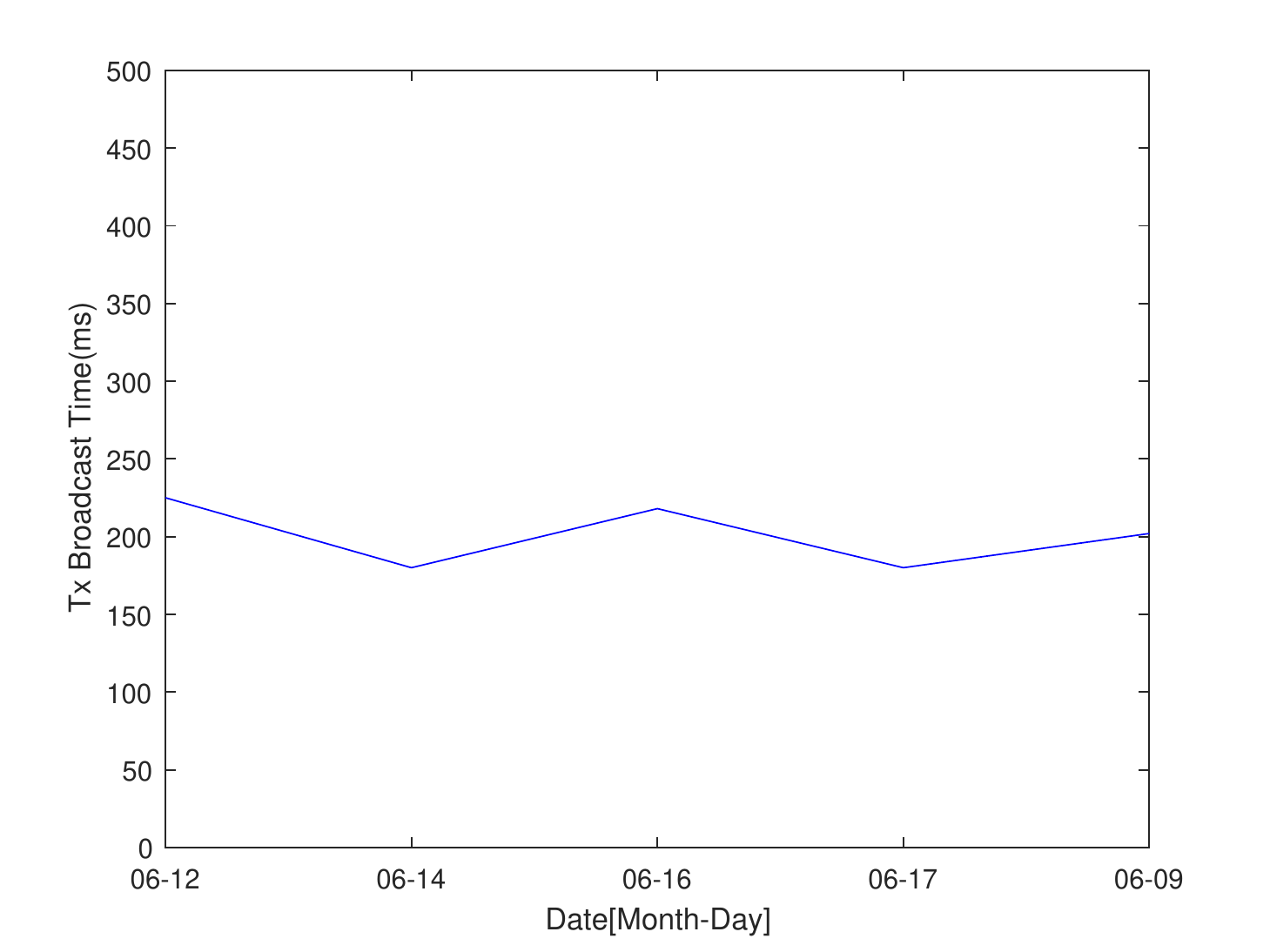}
	\caption{The measured transaction broadcast latency in the Ethereum P2P network.}
	\label{p2p_module}
\end{figure}

\subsection[C]{The analyzing method for the number of hops required to broadcast messages}

In this part, we propose a model for the transactions and blocks broadcast delays in the Ethereum P2P network and use the proposed model and the measured results about the message broadcast delays to analyze the number of hops required to broadcast transactions and blocks over the Ethereum P2P network. The delays of broadcasting a transaction and a block to most of the Ethereum nodes can be mathematically modeled as


\begin{equation}
\begin{split}
{T_{{\rm{BlockDelay}}}}& = {P_{\rm HashBlock}}x({T_{\rm GetHeader}}
 + {T_{\rm GetBody}} \\
&+ {T_{\rm Process}}
 5y) + (1 - {P_{\rm HashBlock}})xy
\end{split}
\end{equation}

\begin{equation}
\begin{split}
{T_{\rm TxDelay}} &= {P_{\rm eth65}}{P_{\rm HashTx}}x({T_{\rm GetHash}} + 3y) \\
&+(1 - {P_{\rm eth65}}{P_{\rm HashTx}})xy
\end{split}
\end{equation}
where the meanings and the used values of the variables are described as below:

\begin{itemize}
\item ${T_{\rm TxDelay}} $ is the average transaction broadcast delay of the Ethereum P2P network. In Section 5.2, we have already found that its value was around 200 ms during the measuring time of June 09, 2020 and June 17, 2020.
\item ${T_{\rm BlockDelay}}$ is the average block broadcast delay of the Ethereum P2P network. Currently, some institutions and teams have measured the average block broadcast time and published the results in real time on the website \cite{amberdata}. We can find from the website \cite{amberdata} that the average block propagation time is 477 ms between June 09, 2020 and June 17, 2020.
\item $x$ is the number of hops required to broadcast transactions and blocks from a neighbor node of {\ttfamily NetworkObserverNode}to to most of the nodes in the Ethereum P2P network.
\item $y$ is the average time to propagate a transaction or a block over one hop. In (2), 5$y$  represents that a node needs 5 times of message propagations to obtains the block after receiving a block hash, as shown in Fig. 4. Similarly, 3$y$  in (3) represents that a node needs 3 times of message propagations to obtains the block after receiving the block hash, as shown in Fig. 6.
\item ${T_{\rm GetHeader}} $ is the waiting time of a node to obtain the block header after receiving the block hash. Its value is 400 ms.
\item ${T_{\rm GetBody}}$ is the waiting time of a node to obtain the block body after receiving the block hash. Its value is 100 ms.
\item ${T_{\rm GetHash}}$ is the waiting time of an eth65 node to obtain a transaction after receiving the transaction hash. Its value is 500 ms.
\item ${T_{\rm Process}}$ is the average time of processing a block at a node. Fig. 13 indicates the block processing time of {\ttfamily LocalFullNode} during the measuring period where it fluctuated around 200 ms. Therefore, we treat the value of the variable ${\rm T_{\rm Process}}$ as 200 ms in our analysis.
\item ${P_{\rm eth65}}$ is the proportion of eth65 nodes in the current network. From June 9, 2020 to June 16, 2020, we observe that there are totally 1380 nodes connected with {\ttfamily NetworkObserverNode} and {\ttfamily LocalFullNode}, of which 40\% were eth65 nodes. Thus, ${P_{\rm eth65}}$ is set to be 0.4.
\item  ${P_{\rm HashBlock}}({P_{\rm HashTx}})$ is the proportion of the blocks(transactions) received by a node after first receiving the hashes of these blocks(transactions) and then requesting these blocks(transactions) to all of the blocks(transactions) received by the node. We have discussed in Section 3 that nodes can receive a block/transaction directly from their neighbor nodes or request a block/transaction after receiving the hash of the block/transaction from their neighbor nodes. The values of ${P_{\rm HashBlock}}$  and ${P_{\rm HashTx}}$ are determined by two factors: i) how long the node will wait after receiving the hash of a block/transaction and before requesting the block/transaction; ii) how many nodes are selected to forward the hash of the block/transaction. As discussed in Section 3, these two factors are the same for the propagations of blocks and transactions. Therefore, we can assume that the values of ${P_{\rm HashBlock}}$  and ${ P_{\rm HashTx}}$ are the same in the Ethereum P2P network. We conducted experiments to measure the value of ${P_{\rm HashBlock}}$ in Ethereum Mainnet during the measuring period of November 2019. We counted the numbers of the blocks received by {\ttfamily LocalFullNode}, and the numbers of the blocks requested by {\ttfamily LocalFullNode} after receiving the block hashes within each day. Fig. 14 presents the measurement results. We can see that indeed there is a part of blocks that are obtained from the block hashes. During the measurement period, the number of blocks obtained from block hashes is 2723, and the number of all the received blocks is 20042, which gives ${P_{\rm HashBlock}=0.135}$. Therefore, we treat the values of ${P_{\rm HashBlock}}$ and  ${P_{\rm HashTx}}$ both as 0.135 in our analysis.
\end{itemize}

So far, we have determined the values of all the variables except that of $x$  and $y$ in (2) and (3). Therefore, after substituting the variable values into (2) and (3), we can solve to obtain $x \approx 3.7$, which means that in average, the broadcast of a block/transaction from one of the neighbor nodes of {\ttfamily NetworkObserverNode} to the whole Ethereum network needs 3.7 hops. This result indicates that the Ethereum P2P network possesses a certain effect of small-world networks \cite{kochen1989small, watts1998collective}, i.e., one node in the network needs no more than 6 hops to reach another node.\footnote{The strict method to analyze whether a P2P network is a small-world network is to construct the topology of the whole P2P network, and then use the topology to compare the degree distribution of the network, the shortest path and the clustering coefficients of this P2P network with that of a random network, as in the method in \cite{humphries2008network}. Since the Ethereum blockchain is a system carrying extremely high economic values, it will inevitably be subject to various attacks. Once some malicious nodes can fully infer the network topology, they can conduct various attacks on the Ethereum P2P network, as some identified attacks on the Bitcoin P2P network \cite{saad2020exploring, heilman2015eclipse}. To prevent these attacks, the design of the Ethereum P2P network intentionally makes it extremely difficult to infer the topology of its P2P network. For example, the Ethereum P2P network protocol does not spread out the timestamps of the communications between the nodes; and the Ethereum P2P network protocol uses K buckets to maintain node information and establish the connections between nodes randomly from the K buckets. As a consequence, the analytical methods used in \cite{humphries2008network} cannot be applied to the Ethereum P2P network. In this work, the measured number of hops required to broadcast messages in the Ethereum P2P network demonstrates that the Ethereum P2P network possesses a certain effect of small-world networks. Strictly verifying whether the Ethereum P2P network is a small-world network is challenging and is left as an open problem.}

\begin{figure}[t]
	\centering
	\includegraphics[width=3.5in]{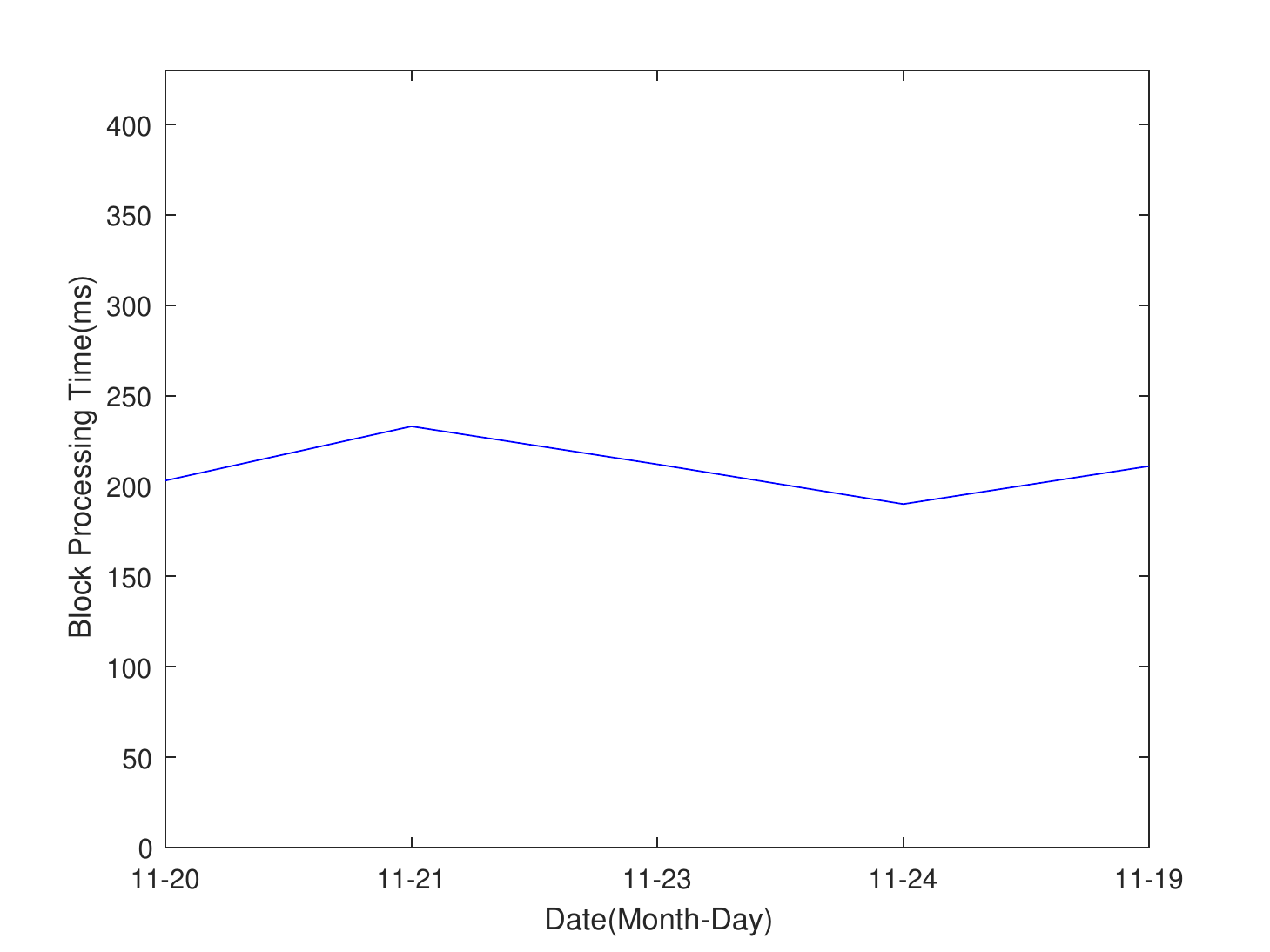}
	\caption{The measured block processing time at {\ttfamily LocalFullNode}.}
	\label{p2p_module}
\end{figure}

\begin{figure}[t]
	\centering
	\includegraphics[width=3.5in]{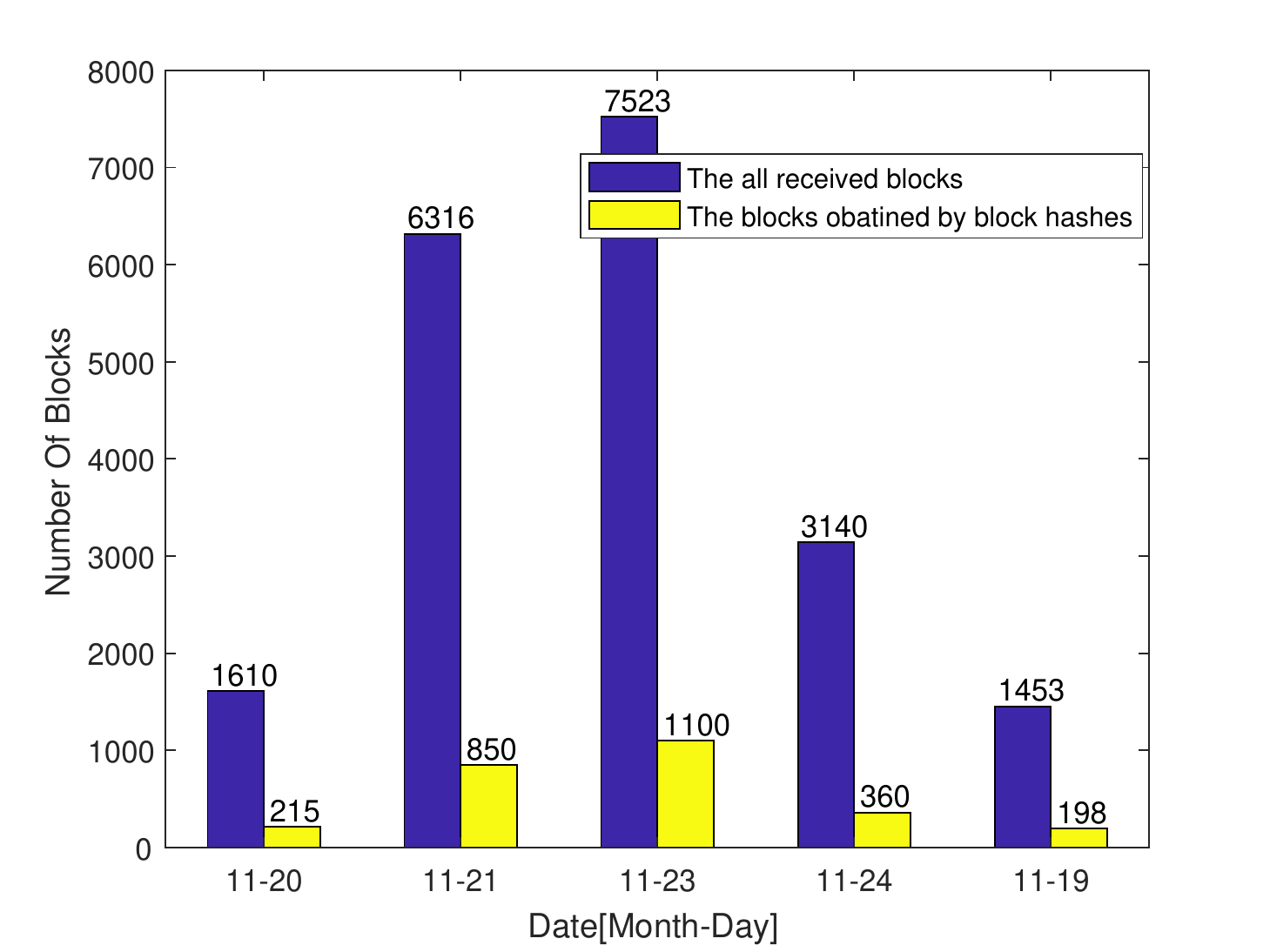}
	\caption{The numbers of all the blocks received by {\ttfamily LocalFullNode} and the numbers of the blocks that are requested by {\ttfamily LocalFullNode} after receiving the block hashes..}
	\label{p2p_module}
\end{figure}

\section{Conclusion}

We proposed Ethna, an Ethereum network analyzer, to probe and analyze the P2P network of the Ethereum blockchain. Ethna sets up probing nodes on Ethereum Mainnet to collects message propagation records and exploits the random feature of the Ethereum message forwarding protocol to analyze the topological characteristics of the Ethereum P2P network. We measured that the average degree of the Ethereum nodes is 47 and there are a few of super nodes with a degree greater than 1000; similar to the P2P network of blockchain systems such as Bitcoin and Monero, the degree distribution of the Ethereum P2P network follows a power-law and has the characteristics of scale-free networks. In addition, we model the message broadcast latencies and analyze the number of hops required to broadcast message over the network with collected message propagation records. We found that messages can be broadcast to most of the Ethereum nodes within 6 hops. This result indicates that there is a small-world effect in the Ethereum P2P network.

\begin{appendix}
	
\noindent {\bfseries1) The reason why we cannot use the transactions forwarded by eth64 nodes to analyze the degrees of eth64 nodes:}	

This appendix first explains why we cannot use the transactions forwarded by eth64 nodes to analyze the degrees of eth64 nodes. When the number of downstream neighbor nodes $N$ for an eth64 nodes is smaller than 16, the formula in (1) does not apply. Each time, an eth64 node will randomly selects $m$ nodes from its $N$ downstream neighbor nodes to forward the block and the value of $m$ is given by
   \begin{equation}
   	\begin{split}
   		m=\left\{
   		\begin{array}{rcl}
   			\sqrt{N }     &      & {N>16}\\
   			4    &      & {4\leq N \leq 16}\\
   			N     &      & {0<N<4}
   		\end{array} \right.
   	\end{split}
   \end{equation}
After that, the eth64 forwards the block hash to the remaining $N - m$ downstream neighbor nodes. As indicated in (4), when $4 \le N \le 16$ for an eth64 node, m is equal to 4 for this range of N. Thus, we cannot know $N$ from $m$. Due to this restriction in the block forwarding strategy of eth64 nodes, we cannot measure the degrees of eth64 nodes with low degrees using the transactions forwarded by eth64 nodes.

\noindent {\bfseries2) The reason why we cannot use the forwarding of blocks to analyze the node degrees for eth65 nodes or eth64 nodes:}	

This appendix then explains why we cannot use the number of block packets ( {\ttfamily BlockPacketCount}) and the number of block hash packets ( {\ttfamily BlockHashPacketCount}) to analyze the node degrees for eth65 nodes or eth64 nodes, i.e., why we cannot replace {\ttfamily TxPacketCount} in (1) with  {\ttfamily BlockPacketCount} and replace {\ttfamily TxHashPacketCount} in (1) with  {\ttfamily BlockHashPacketCount} to compute the number of neighbor downstream nodes for each eth65 or eth64 node. The numbers of block packets and block hash packets sent by eth64 nodes or eth65 nodes are too small during the measuring period. According to the website \cite{Etherscan}, currently there are 30-50 new transactions generated per second in Ethereum, and one block generated in every 15 seconds. If {\ttfamily NetworkObserverNode} or {\ttfamily LocalFullNode} keeps a stable connection with a node for 1 hour, {\ttfamily NetworkObserverNode} or {\ttfamily LocalFullNode} can receive approximately 240 block and block hash packets from this node, and {\ttfamily NetworkObserverNode} or {\ttfamily LocalFullNode} can receive about 108000 transaction and transaction hash packets from this node. Moreover, the {\ttfamily NetworkObserverNode} or {\ttfamily LocalFullNode} usually connect with each node for less than 1 hour, so the value of  {\ttfamily BlockPacketCount} and the value of  {\ttfamily BlockHashPacketCount} for each node are even smaller than 240. Since the formula in (1) is hold only when a large number of messages are randomly forwarded to ensure the statistical property, it is inaccurate to use  {\ttfamily BlockPacketCount} and  {\ttfamily BlockHashPacketCount} to measure the degrees of nodes.

\end{appendix}

\ifCLASSOPTIONcaptionsoff
  \newpage
\fi

\bibliographystyle{IEEEtran}

\bibliography{refs}

\end{document}